\documentclass{JHEP3}

\usepackage[normalem]{ulem}
\usepackage{cite}
\usepackage{epsfig}
\usepackage{amsmath}
\usepackage{bm,longtable,enumerate}
\def\beq{\begin{equation}}
\def\eeq{\end{equation}}
\def\beqa{\begin{eqnarray}}
\def\eeqa{\end{eqnarray}}
\def\ss{\scriptscriptstyle}

\title{Black-hole quasinormal modes and scalar glueballs
in a finite-temperature AdS/QCD model}

\author{Alex S. Miranda$^{\dag,a}$, C. A. Ballon Bayona$^{\ddag}$, 
Henrique Boschi-Filho$^{\dag,b}$, and Nelson R. F. Braga$^{\dag,c}$
\\
$^{\dag}$Instituto de F\'{i}sica, Universidade
Federal do Rio de Janeiro,\\
Caixa Postal 68528, RJ 21941-972, Brazil\\
$^{a}$email: astmiranda@if.ufrj.br;
$^{b}$email: boschi@if.ufrj.br;
$^{c}$email: braga@if.ufrj.br\\
\\
$^\ddag$Centro Brasileiro de Pesquisas F\'{i}sicas,
Rua Dr. Xavier Sigaud 150,\\
Urca, 22290-180, Rio de Janeiro, RJ, Brazil\\
email: ballon@cbpf.br}

\abstract{
We use the holographic AdS/QCD soft-wall model to investigate
the spectrum of scalar glueballs in a finite temperature plasma.
In this model, glueballs are described by a massless scalar field in an 
$\mbox{AdS}_{5}$ black hole with a dilaton soft-wall background. 
Using AdS/CFT prescriptions, we compute the boundary 
retarded Green's function. The corresponding thermal spectral function 
shows quasiparticle peaks at low temperatures. 
We also compute the quasinormal modes of the scalar field in the soft-wall 
black hole geometry. The temperature and momentum dependences of these modes 
are analyzed. The positions and widths of the peaks of the spectral
function are related to the frequencies of the quasinormal modes.
Our numerical results are found employing the power series method 
and the computation of Breit-Wigner resonances.}


\preprint{}

\begin{document}


\section{Introduction}
\label{introd}

In the last years the AdS/CFT correspondence \cite{Maldacena:1997re}
has been recognized as an important tool to investigate
diverse aspects of strongly coupled gauge theories. This correspondence
relates a ten-dimensional string theory or eleven-dimensional M-theory
to a conformal gauge theory on the corresponding spacetime boundary.
The most celebrated version of this correspondence relates string
theory in $\mbox{AdS}_{5}\times S^{5}$ space to four-dimensional
$\mbox{SU}(N_{c})$ Yang-Mills gauge theory with large $N_{c}$
and extended ${\cal{N}}=4$ supersymmetry.

At low energy string theory is represented by an effective
supergravity theory. Hence, the AdS/CFT correspondence implies
a gauge/gravity duality. Using this duality, one can calculate
quantities like quantum correlation functions for gauge-field
boundary operators from the classical supergravity action
for bulk fields \cite{Witten:1998qj,Gubser:1998bc}. 

One can formulate a finite temperature version of the AdS/CFT correspondence 
including a black hole in the AdS spacetime.
The Hawking temperature of the black hole, related to the horizon
position, is identified with the temperature of the dual gauge theory
\cite{Witten:1998zw}. Following this approach, one can calculate various
thermal Green's functions for the boundary gauge theory by taking
derivatives of the on-shell bulk action for fields subjected to certain
conditions. In the Euclidean version of the AdS/CFT correspondence,
the prescription to compute Green's functions is the same found in
\cite{Witten:1998qj,Gubser:1998bc} with the classical bulk fields
uniquely determined by their value at the boundary and the requirement
of regularity at the horizon. In Minkowski space, a prescription for
calculating retarded propagators at finite temperature was proposed in
\cite{Son:2002sd} (see also Refs. \cite{Herzog:2002pc,Skenderis:2008dh,
Skenderis:2008dg}) imposing that the fields satisfy purely incoming-wave
condition at the horizon. Physically, this means that the black holes
only absorb without any emission. An important consequence of this
prescription is that the poles of the retarded correlation function of
an operator in the boundary conformal field theory correspond to the
quasinormal (QN) frequencies of the holographic dual field in the
AdS black-hole spacetime. In the context of the AdS/CFT correspondence,
quasinormal modes have been object of study of many works 
(see Ref. \cite{Berti:2009kk} for a recent review).

The AdS/CFT correspondence can be also an
important tool to study many problems in strong interactions,
like the calculation of hadronic spectra.
The theory that describes strong interactions is QCD,  a $\mbox{SU}(3)$ Yang-Mills 
theory with a coupling constant that runs with the energy.
At low energies the coupling is strong and one can not use perturbative methods. 
In this regime, one needs other approaches like lattice QCD or phenomenological 
effective models. Recently, the idea of modifying the AdS space introducing an
infrared cut off led to  phenomenological models to describe strong interactions.
This modification implies a mass gap in the boundary gauge theory that breaks the conformal symmetry  
present in the AdS/CFT correspondence. This kind of approach is known as AdS/QCD.

The simplest AdS/QCD model, known as hard wall, introduces
a mass gap via a cut off on the bulk geometry. In this way it
is possible to use the gauge/gravity duality to estimate hadronic
masses \cite{BoschiFilho:2002ta,BoschiFilho:2002vd,deTeramond:2005su,
Erlich:2005qh,DaRold:2005zs,BoschiFilho:2005yh}. The hard wall is
confining at low temperatures \cite{BoschiFilho:2006pe} and
deconfining at high temperatures \cite{BoschiFilho:2005mw}. Another
simple model of AdS/QCD, known as soft-wall, introduces a mass gap
through a non-uniform background scalar field in the AdS spacetime.
This soft-wall model presents linear Regge trajectories for vector 
mesons and glueballs \cite{Karch:2006pv,Colangelo:2007pt,Colangelo:2008us}. 
Confinement/deconfinement thermal phase transition in these two models 
was studied in \cite{Andreev:2006eh,Herzog:2006ra,Kajantie:2006hv,BallonBayona:2007vp}. 
It was found in \cite{Herzog:2006ra} that this corresponds to a gravitational 
Hawking-Page phase transition \cite{Hawking:1982dh} between spaces with and without a black hole. 
This is a first order phase transition, where two different phases can coexist. 
The black-hole spacetime is thermally favored at high temperatures, while at temperatures
below some critical model-dependent value, the thermally favored space is a pure AdS.

In this article we study the spectrum of scalar glueballs 
at finite temperature using the soft-wall AdS/QCD model.
This model, at high temperatures, consists of an AdS space with a nonuniform
background dilaton field and a black hole. At intermediate and low
temperatures, we consider, respectively, the (supercooled) metastable and
unstable phases of the plasma where the black hole is still 
present\footnote{A similar situation was considered recently for the case of 
mesons in the D3-D7 model \cite{Myers:2007we}.}. 

Our motivation for studying the supercooled phase of the plasma is to gain some insight
into the mechanism of glueball formation when the plasma cools down. We know that at the
high temperatures reached when the plasma is formed, there are no stable hadronic states. 
On the other hand, at zero temperature, when the plasma disappears, there should be only 
colorless hadronic states with a well defined mass spectrum. 
These two separate regimes have been extensively studied using gauge/gravity duality. 
We want to investigate in this article how the transition
between the high temperature deconfined phase and the zero temperature confined phase
takes place. We think that this kind of analysis may be
useful to understand hadronic formation in heavy ion  collisions.

It is important to remark that the five dimensional background of the soft-wall 
phenomenological model does not arise as a solution of the 5D Einstein equations. 
In this model there is a time-independent background scalar field $\Phi$ 
that is a function of an infrared energy scale. In this article we will
study the dynamics of an extra  scalar field $\phi$ in this background
at finite temperature. As the infrared energy scale goes to zero, 
the background $\Phi$ of the soft-wall model vanishes. 
In this limit, we recover the usual AdS/CFT picture and the scalar
field $\phi$ can be thought as a perturbation of the trivial dilaton vacuum
or of the scalar part of the metric. In the AdS/CFT correspondence this field 
is dual to the scalar glueball operator. Motivated by this fact, we assume
in this work that the field $\phi$ is dual to scalar glueballs even in the  
presence of the infrared scale.

Therefore, we identify the quasinormal modes (QNM)
of the massless scalar field in the bulk as the spectrum of scalar glueballs
in the boundary. We study the zero- and finite-temperature retarded Green's
functions of scalar glueball operators both with numerical and analytical
techniques. The spectral function is identified with the imaginary part of the 
retarded Green's function. We perform also a numerical analysis to obtain the real and imaginary parts of the
quasinormal frequencies as functions of the temperature and the spatial
momentum. In such analysis a good convergence is achieved using the
power series method \cite{Horowitz:1999jd} at high temperatures and
the Breit-Wigner ressonance method \cite{Berti:2009wx} at low temperatures.


\section{Glueballs in the soft-wall model at zero and finite temperature}
\label{model}

In this section we obtain the equation of motion of a scalar field in the
soft-wall model at zero and finite temperature and transform this equation
into a Schr\"odinger-like form. We also describe the wave functions that
are relevant for the computation of the glueball retarded Green's function
and analyze the corresponding effective potential.

\subsection{The soft-wall model at zero temperature}

According to the AdS/CFT correspondence, the particle states on the boundary
field theory are dual to normalizable fields in the bulk spacetime. As a
consequence of the conformal symmetry, the mass spectrum of these states is
continous. 
In the AdS/QCD models the conformal invariance is broken by the presence of an
infrared cut-off turning the particle mass spectrum discrete.  The soft-wall
model of Ref. \cite{Karch:2006pv} was originally formulated at zero temperature
in order to reproduce the approximate linear Regge trajectories for vector
mesons. It consists of a five-dimensional AdS spacetime and a background
dilaton-like field $\Phi(z)$. The $\mbox{AdS}_{5}$ metric in Poincar\'e coordinates
is given by 
\begin{equation}
ds^{2}=e^{2A(z)}\left[-dt^2+\sum_{i=1}^{3}(dx^{i})^{2}+dz^2\right],
\label{AdS}
\end{equation}
where $A(z)=-\mbox{ln}(z/L)$ and $L$ denotes the AdS curvature radius.
The interaction of the background field $\Phi(z)$ with the bulk fields
is defined through the replacement of the action integrals: 
\begin{equation}
\int d^{5}x\sqrt{-g}\,{\cal L} \qquad\Longrightarrow \qquad
\int d^{5}x\sqrt{-g}e^{-\Phi}{\cal L}\;.
\label{action00}
\end{equation}

\noindent The factor $\exp[-\Phi(z)]$ with $\Phi(z)=cz^2$ introduces an
effective infrared cut-off in the AdS space, with $\sqrt{c}\,$ playing
the role of a mass scale. This modifies the normalizability condition for
bosonic fields in the bulk. The solutions with plane wave dependence
$\exp(ik\cdot x)$ become normalizable only for particular discrete values
of $k^{2}=-m_{n}^{2}$, with nonvanishing $m_n$ and $n=0,1,2,...\,$. This
implies a mass gap in the dual gauge theory. For instance, the spectrum
of vector mesons is \cite{Karch:2006pv}
\begin{equation}
m_{{\ss{V}}_{n}}^{2}=4c(n + 1) \, ,
\end{equation}
where $n$ is the radial quantum number.

Scalar glueball states at zero temperature in the soft-wall model
were studied in \cite{Colangelo:2007pt} by considering
a massless scalar field living in the AdS spacetime \eqref{AdS}
with the same dilaton background field $\Phi(z)\,=\,cz^2$.
The corresponding glueball spectrum is 
\begin{equation}
m_{{\ss{G}}_{n}}^{2}=4c(n+2)\,.
\label{T0spectrum}
\end{equation}
The parameter $\sqrt{c}\,\,$, in this case, sets the scale of the glueball masses.
Other recent results on glueballs in holographic models can be found in \cite{Forkel:2007ru}. 
For a recent review on glueballs see \cite{Mathieu:2008me}.

\subsection{The finite temperature case} 

The soft-wall model at finite temperature is composed by the dilaton
field $\Phi=cz^2$ and an AdS black-hole spacetime with
translationally invariant horizon: 
\begin{equation}
ds^{2}=e^{2A(z)}\left[-f(z)dt^{2}+\sum_{i=1}^{3}(dx^{i})^{2}+
f(z)^{-1}dz^2\right],
\label{background}
\end{equation}
where $f(z)=1-(z/z_{h})^4$. The coordinate $z$ is defined in the range
$0\le z \le z_h$ and $z=0$ corresponds to the boundary of the space. 
The parameter $z_{h}$ indicates the position of the event horizon,
which is related to the black-hole Hawking temperature $T$ by
$z_{h}=1/\pi T$. According to the holographic dictionary, $T$ also
represents the temperature of the boundary field theory.

\subsubsection*{Thermodynamics}

Considering the Euclidean version  of the metrics given by \eqref{AdS}
and \eqref{background}  with compactified times, 
Herzog \cite{Herzog:2006ra} has shown that,
in the presence of the dilaton field $\Phi=cz^2$, there is a thermal
competition between these spaces, which is characterized by the ratio
$\widetilde{T}=\pi T/\sqrt{c}$. The regularized gravity action densities for the 
thermal AdS and AdS black hole spaces are given by \cite{BallonBayona:2007vp}
\beq
S_{\ss{AdS}} \,=\,  \frac{N_c^2}{4\pi^2} \left[ c^2
\left( \frac32 - \gamma\right) \beta \right] \,,
\eeq
\beq
S_{\ss{BH}} \,=\, S_{\ss{AdS}} \,+\, \frac{N_c^2}{4\pi} \left[ c^2 z_h Ei (-cz_h^2) 
+ e^{-cz_h^2} \left(\frac{c}{z_h} - \frac{1}{z_h^3}\right) + \frac{1}{2z_h^3} \right] \,,
\label{BHgaction}
\eeq
where $Ei(u) = -\int_{-u}^{\infty} dt \, t^{-1} e^{-t}$ and $z_h=\beta/\pi$. 
At low temperatures the thermal AdS space is thermodynamically preferred 
compared to the AdS black-hole space, 
since it has a smaller gravity action.
Then a first-order Hawking-Page type phase transition between these two
spaces occurs at $\widetilde{T}_{c}^{\,2} \approx 2.38644$. This transition
is interpreted in the boundary field
theory as a confinement/deconfinement phase transition as can be seen
by calculating the corresponding entropy change.
At $T>T_{c}$ the AdS black-hole space has a smaller gravity action 
so it is the dominant configuration.
This space is dual to the plasma phase of a strongly coupled Yang-Mills
theory with large number of colors. In the left panel of Fig. \ref{transicao}, 
we show the difference of the regularized gravity actions densities,
$S_{\ss{BH}}-S_{\ss{AdS}}$, that determines the confinement/deconfinement
phase transition.

\FIGURE{
\centering\epsfig{file=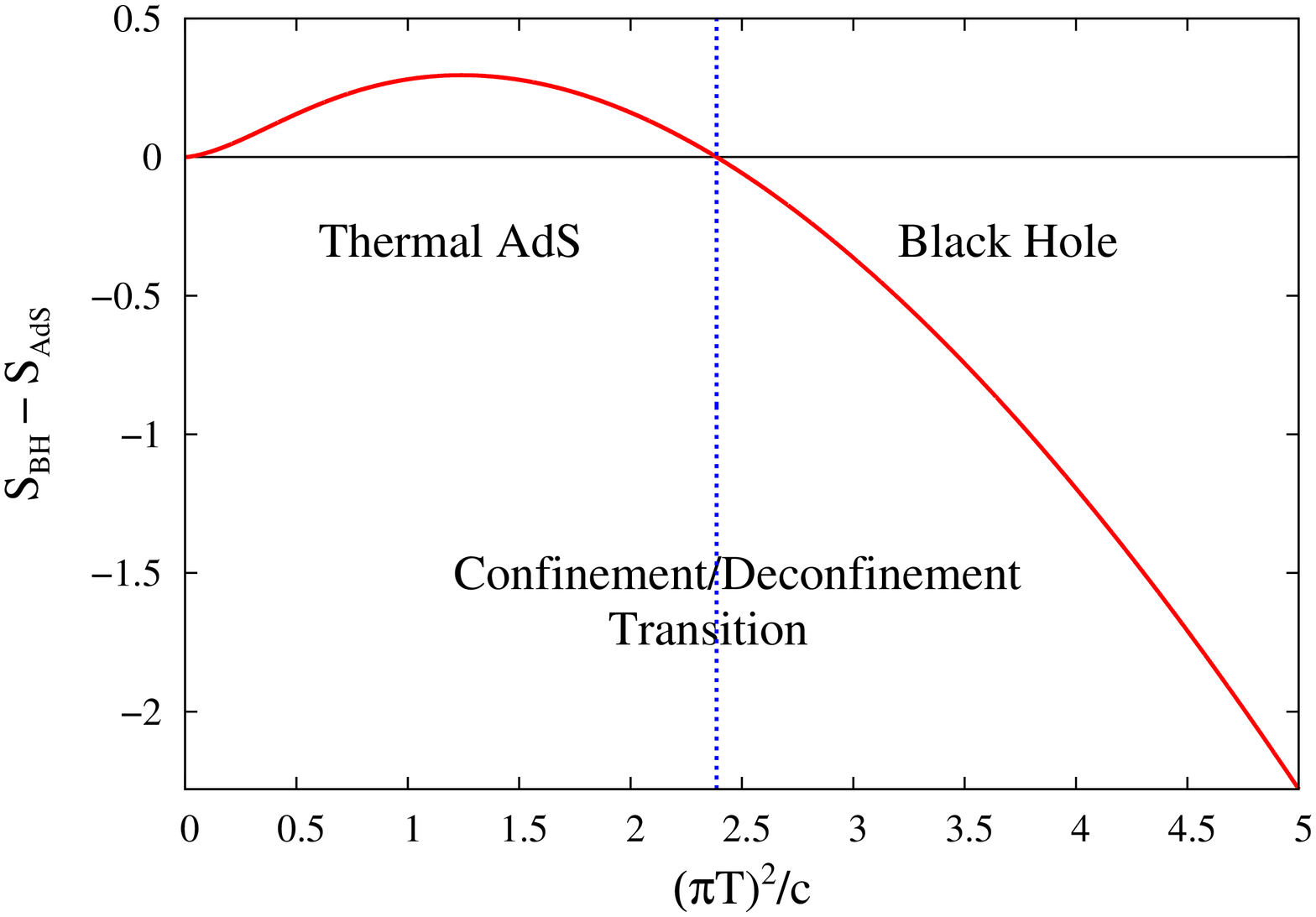,
width=5.15cm, angle=270}
\centering\epsfig{file=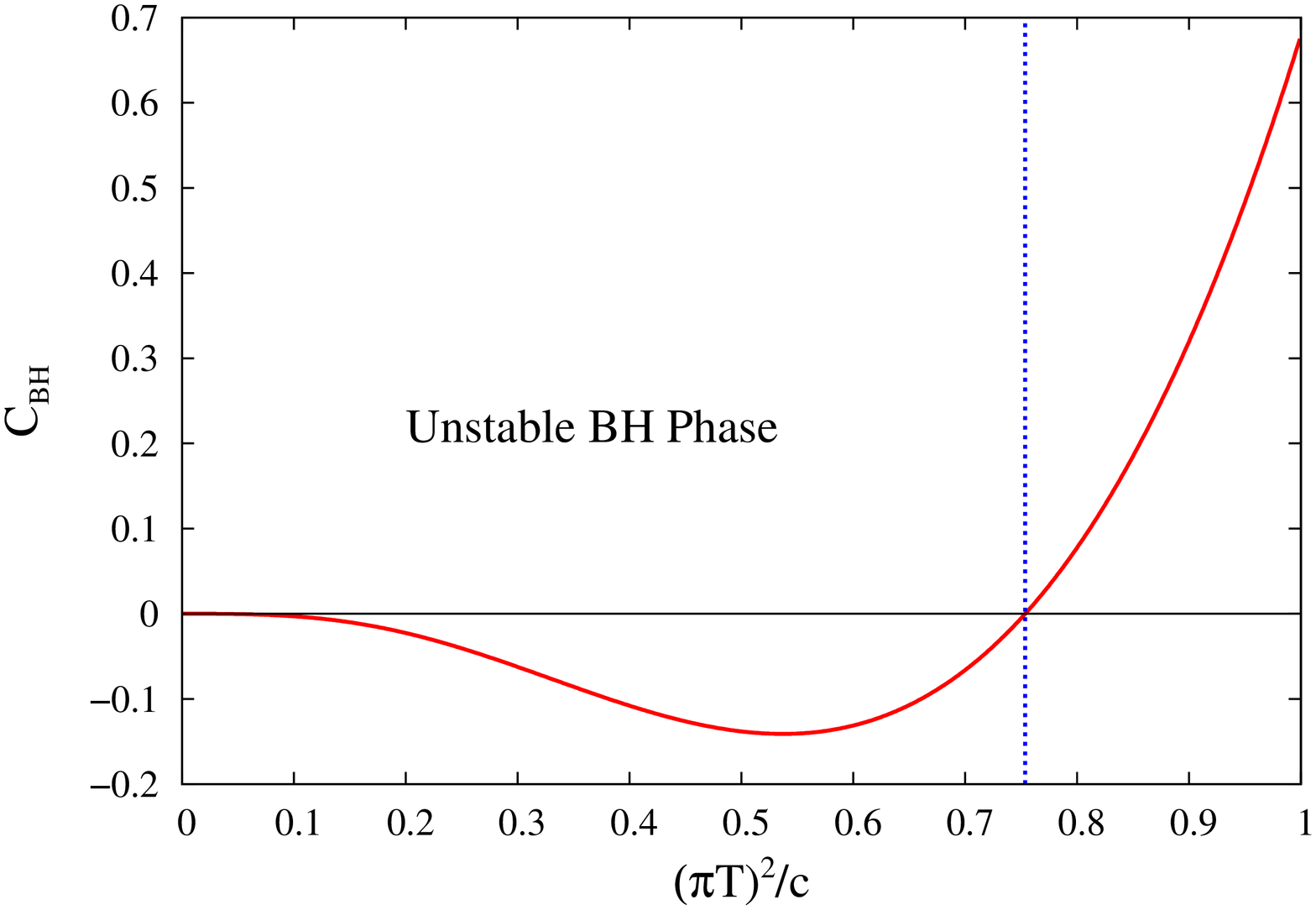,
width=5.15cm, angle=270}
\caption{{\sl Left}: Difference of regularized gravity action densities 
in the soft-wall model as a function of $(\pi T)^2/c$. 
{\sl Right}: Specific heat density for the AdS black hole with
the soft-wall dilaton as a function of $(\pi T)^2/c$. The quantities
$S_{\ss{BH}}-S_{\ss{AdS}}$ and ${\cal C}_{\ss{BH}}$ are in units
of $N_c^2 c^{\,3/2}/4 \pi$.}
\label{transicao}}

In this article, we consider the black-hole spacetime \eqref{background}
for all temperatures. For temperatures lower than $T_{c}$ the black hole
is not stable and can be interpreted as a supercooled Yang-Mills plasma. 
If the sign of the specific heat is positive (negative) it will be in a
metastable (unstable) phase. From the regularized action density 
for the soft-wall AdS black hole \eqref{BHgaction}, we obtain the
specific heat density 
\beqa
{\cal C}_{\ss{BH}} \,\equiv\, - \beta^2 \frac{\partial^2}{\partial\beta^2}
S_{\ss{BH}}  \,=\, \frac{\pi^2}{2} \, N_c^2 T^3 
\Big  \{ e^{-\frac{c}{(\pi T)^2}} \left[ \frac{4 c}{(\pi T)^2}
+ 6 \right] - 3 \Big  \}  \, . 
\eeqa
In the limit $c \to 0$ we recover the usual (positive) result
for the black hole without dilaton. In our case, the presence of
the dilaton implies that the specific heat  is negative for
$\widetilde{T}^2  \lesssim 0.75$, as shown in the right panel
of Fig. \ref{transicao}.

\subsubsection*{Equation of motion}

The dual of the scalar glueball operator $\mbox{Tr}(F^2)$ is a
massless scalar field $\phi$ in the bulk. In the finite temperature
soft-wall model this field is described by the action
\begin{equation}
S=-\frac{\pi^3 L^5}{4\kappa_{10}^2}\int d^{5}x\sqrt{-g}
e^{-\Phi}\,g^{\ss{MN}}\partial_{\ss{M}}\phi\partial_{\ss{N}}\phi \, ,
\label{action0}
\end{equation}
where $\Phi(z)=cz^2$ is again the soft-wall dilaton field and $g_{\ss{MN}}$
is the black-hole metric given by \eqref{background}. The constants $L$
and $\kappa_{10}$ are the anti-de Sitter radius and the ten-dimensional
gravitational constant. The indices $M,N$ run over $0,\,1,\,...,4$ where 
$x^{\mu}$ ($\mu=0,...,3$) are the 4$d$ boundary coordinates and $x^{4}=z$
is the extra radial coordinate.

As usual in the soft-wall model, we do not consider here the backreaction of
the dilaton $\Phi$ on the metric. From the action \eqref{action0}, we find
the scalar field $\phi$ satisfies the equation
\begin{equation}
\frac{e^{\Phi}}{\sqrt{-g}}\partial_{z}\left(\sqrt{-g}e^{-\Phi}
g^{zz}\partial_{z}\phi\right)+g^{\mu\nu}\partial_{\mu}\partial_{\nu}\phi\,=\,0
\, .\label{eqwave1}
\end{equation}
Introducing the Fourier transform with respect to the coordinates $x^{\mu}$,
\begin{equation}
\phi(z,x)=\int\frac{d^{4}k}{(2\pi)^4}e^{ik \cdot x}\bar\phi(z,k) \,,
\end{equation}
equation \eqref{eqwave1} takes the form
\begin{equation}
e^{B}f(z)\partial_{z}\left(e^{-B}f(z)\partial_{z}\bar{\phi}\right)+
\left(\omega^{2}-f(z)q^{2}\right)\bar{\phi}=0 \, ,
\label{eqwave2}
\end{equation}
where $k_{\mu}=(-\omega,q_{i})$, $\,q^{2}=\sum_{i=1}^{3}q_{i}^{2}\,$
and $\,B=\Phi-3A=cz^2+3\,\mbox{ln}(z/L)$.

\subsection{The Schr\"odinger equation and the wave functions}
\label{wave1}

In the study of classical-field perturbations around black holes, it is
usually convenient to introduce the Regge-Wheeler tortoise coordinate
$r_{\ast}$. This coordinate is defined by the relation
$\partial_{r_{\ast}}=\,-f(z) \partial_z$ with $z$ in the
interval $0 \le z \le z_{h}$. Since in our
case $1/f(z)$ is integrable we find the expression 
\begin{equation}
r_{\ast}=\frac{1}{2}\,z_{h}\left[-\arctan{\left(\frac{z}{z_{h}}\right)}+
\frac{1}{2}\ln{\left(\frac{z_{h}-z}{z_{h}+z}\right)}\right].
\label{rtart}
\end{equation}
Note that the horizon $z=z_{h}$ corresponds to $r_{\ast}\to -\infty$ and the
boundary $z=0$ is chosen at $r_{\ast}=0$. It is also convenient to use
a Bogoliubov transformation of the form $\psi=e^{-B/2}\bar\phi$. Then,
Eq. \eqref{eqwave2} reduces to a one-dimensional Schr\"odinger equation:
\begin{equation}
\partial_{r_{\ast}}^{2}\psi+\omega^2 \psi=V\psi,
\label{Schrod1}
\end{equation}
where the effective potential $V$ is given by
\begin{equation}
V=f(z)q^2+e^{B/2}\partial_{r_{\ast}}^{2}e^{-B/2}. 
\end{equation}
This potential can be explicitly written in terms of the coordinate $z$ as 
\begin{equation}
V(z)=\frac{f(z)}{z^2}\left[q^{2}z^{2}+\frac{15}{4}+\frac{9}{4}
\frac{z^{4}}{z_{h}^4}+2cz^{2}\left(1+\frac{z^4}{z_h^4}\right)+
c^{2}z^{4}f(z)\right].
\label{potential}
\end{equation}

By investigating the asymptotic behavior of the Schr\"odinger
equation \eqref{Schrod1} near the boundary $z=0$ we find two particular
real solutions $\psi_{1}$ and $\psi_{2}$. These solutions have
the following asymptotic form:
\begin{gather}
\psi_{1}=z^{5/2}\left[1+a_{11}z^{2}+a_{12}z^{4}+\cdots\right],
\label{normalizable}\\ 
\psi_{2}=z^{-3/2}\left[1+a_{21}z^{2}+a_{22}z^{4}+\cdots\right]+
b\,\psi_{1}\,\mbox{ln}(cz^2),
\label{nonnormalizable}
\end{gather}
where the ellipses denote higher powers of $z$ and the
coefficients above are given by
\begin{equation}
\begin{aligned}[3]
&a_{11}=-\frac{\omega^2-q^2-2c}{12},\quad & a_{12}=&
\frac{(\omega^2-q^2-2c)^2}{12 \times 32}+\frac{1}{2z_{h}^{4}}+
\frac{c^{2}}{32},\\
&a_{21}=\frac{\omega^2-q^2-2c}{4},\quad & b=&-\frac{(\omega^2-q^2)
(\omega^2-q^2-4c)}{32}.
\label{coeffsol}
\end{aligned}
\end{equation}
The wave function $\psi_{1}$ is  {\textit{normalizable}} due to
its asymptotic behavior $z^{5/2}$, while the wave function $\psi_{2}$
is {\textit{non-normalizable}} since it behaves asymptotically as
$z^{-3/2}$. The coefficient $a_{22}$ is arbitrary and, in particular,
we can choose $a_{22}=0$. The solutions $\psi_{1}$ and $\psi_{2}$ will
be useful to construct the glueball retarded Green's function in section
\ref{Greenfunction}, and to compute the black-hole quasinormal modes
in section \ref{spectrum}. 

Another special kind of solutions to the Schr\"odinger equation
\eqref{Schrod1} are the {\textit{ingoing}} and {\textit{outgoing}}
wave functions, $\psi_{-}$ and $\psi_{+}$. These solutions are
motivated by the fact that the potential $V(z)$ vanishes when
$z \to z_{h}$ so that Eq. \eqref{Schrod1} takes the form of a 
free particle equation with solutions proportional to
$\exp(\pm i\omega r_{\ast})$. Since in our convention for the
Fourier transform the temporal dependence of $\bar{\phi}$ is
$\exp(-i\omega t)$, the minus (plus) sign can be interpreted as an
ingoing (outgoing) plane wave in the sense that the wave travel into
(out of) the horizon (localized at $r_{\ast} \to -\infty$)
as the time increases. The solutions $\psi_{\pm}$ can be expressed
in terms of the normalizable and non-normalizable solutions:
\begin{equation}
\psi_{\pm}={\cal A}_{{\ss{(\pm)}}}\psi_{2}+{\cal B}_{\ss{(\pm)}}
\psi_{1}\, ,\label{ingoingoutgoingexp}
\end{equation}
where ${\cal A}_{{\ss{(\pm)}}}$ and ${\cal B}_{\ss{(\pm)}}$ are
connection coefficients associated to Eq. \eqref{Schrod1},
and determined as functions of $\omega$ and $q$ by imposing the boundary
conditions on the field solution.

The Schr\"odinger equation \eqref{Schrod1} can be expanded near the
event horizon leading to the following expansion for $\psi_{-}$ and $\psi_{+}$: 
\begin{equation}
\label{nearhorizonexp}
\psi_{\pm}=e^{\pm i \omega r_{\ast}}\left[1+a_{1{\ss{(\pm)}}}
\left(1-\frac{z}{z_h}\right)+a_{2{\ss{(\pm)}}}\left(1-\frac{z}{z_h}
\right)^{2}+\cdots\right],
\end{equation}
where the ellipses denote higher powers of $(1-z/z_{h})$ and
the coefficients appearing in the foregoing expression are
\begin{equation}
\begin{split}
a_{1{\ss{(\pm)}}}=& \frac{6+(q^{2}+4c)z_{h}^{2}}{4\pm 2i\omega z_{h}},\\
a_{2{\ss{(\pm)}}}=& \frac{1}{16\pm 4i\omega z_{h}}
\Big\{\left[22\pm 4i\omega z_{h}+(q^{2}+4c)z_{h}^{2}\right]
a_{1{\ss{(\pm)}}}-2z_{h}^{2}(q^{2}+8c)-9+4c^{2}z_{h}^{4}\Big\}.
\end{split}
\label{coefa1a2}
\end{equation}

The ingoing and outgoing wave functions form a basis for any other
wave function. In particular, the normalizable and non-normalizable
solutions $\psi_{1}$ and $\psi_{2}$ can be expressed as 
\begin{equation}
\psi_{2}={\cal C}_{\ss{(2)}}\psi_{-}+{\cal D}_{\ss{(2)}}\psi_{+}\,,
\qquad\psi_{1}={\cal C}_{\ss{(1)}}\psi_{-}+{\cal D}_{\ss{(1)}}\psi_{+}\,.
\label{inverserelation}
\end{equation}
From (\ref{ingoingoutgoingexp}) and (\ref{inverserelation})
we find a useful relation between the connection coefficients 
\beqa
\begin{pmatrix}    
{\cal A}_{\ss{(-)}} & \;{\cal B}_{\ss{(-)}} \\
{\cal A}_{\ss{(+)}} & \;{\cal B}_{\ss{(+)}}
\end{pmatrix} \,=\,  \begin{pmatrix}    
{\cal C}_{\ss{(2)}} & \;{\cal D}_{\ss{(2)}} \\
{\cal C}_{\ss{(1)}} & \;{\cal D}_{\ss{(1)}} 
\end{pmatrix}^{-1}. \label{coeffrelation}
\eeqa

At zero temperature, it is not possible to impose the ingoing
(outgoing) condition at the ``horizon'' $z=\infty$ due to the
presence of the background dilaton field. Instead, we may impose
the regularity condition
\begin{equation}
\lim_{z \to \infty}\psi_{0}=0\,,
\label{regularity}
\end{equation}
where $\psi_{0}$ is a solution of Eq. \eqref{Schrod1} at $T=0$.
This wave function can be decomposed as 
\begin{equation}
\psi_{0}={\cal A}_{\ss{(0)}}\psi_{2}+{\cal B}_{\ss{(0)}}\psi_{1},
\end{equation}
where ${\cal A}_{\ss{(0)}}$ and ${\cal B}_{\ss{(0)}}$ are the
zero-temperature connection coefficients associated to \eqref{Schrod1}.

\subsection{An analysis of the effective potential}

We analyze here the effective potential $V(z)$, given explicitely by
\eqref{potential}, as a function of $r_{\ast}$. At zero temperature
($z_{h}\to \infty$), the tortoise coordinate $r_{\ast}$ reduces to $-z$
and the potential takes the form
\begin{equation}
V_{\ss{\,T=0\,}}(z)=\frac{1}{z^2}\left[\frac{15}{4}+
(q^{2}+2c)z^{2}+c^{2}z^{4}\right]\, ,
\label{potentialzerotemp}
\end{equation}
which is the potential considered in Ref. \cite{Colangelo:2007pt},
and that leads to the glueball spectrum of Eq. \eqref{T0spectrum}.
In a similar way, the limit $c\to 0$ reduces the effective potential to 
\begin{equation}
V_{\ss{\,c=0\,}}(z)=\frac{f(z)}{z^2}\left[q^{2}z^{2}+\frac{15}{4}+\frac{9}{4}
\frac{z^4}{z_h^4}\right]\, ,
\label{nodilatonpotentialVz}
\end{equation}
which is the potential obtained in Refs. \cite{Gibbons:2002pq,Kodama:2003jz,
Morgan:2009pn} in the absence of the dilaton. The effective  potential $V(z)$
suffers a very interesting transition in shape when going from high to low
temperatures. Such a transition is discussed below.

\FIGURE{
\centering\epsfig{file=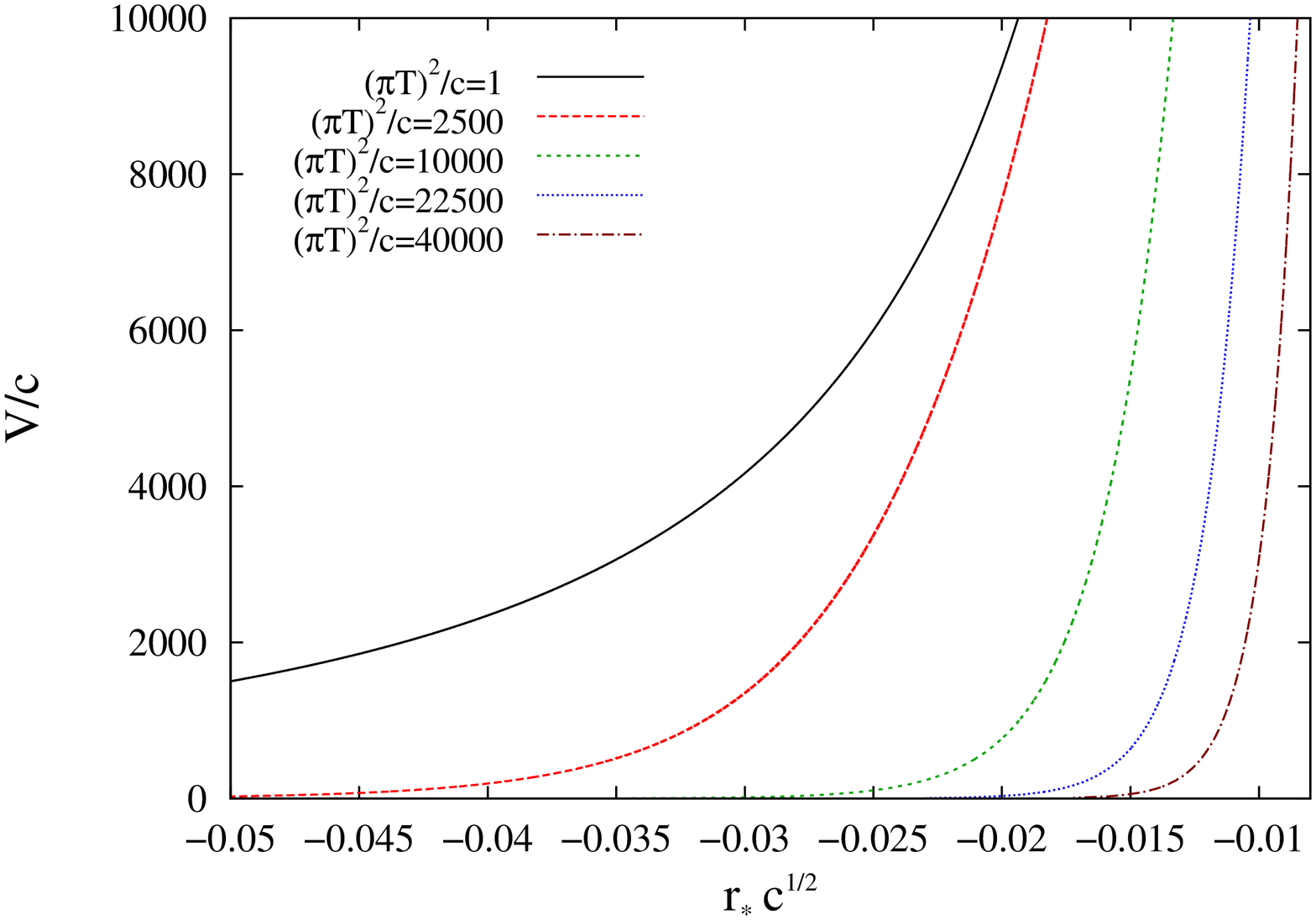,
width=7.2cm, angle=270}
\caption{The behavior of the effective potential $V$ for
high- and intermediate-temperature values.}
\label{potencial2}}

\subsubsection*{High temperatures}

In Fig. \ref{potencial2}, we show the potential as a function of the tortoise
coordinate in the high- and intermediate-temperature regimes for $q=0$.
We observe the absence of any potential well so that there are no bound
states in these regimes. For any fixed value of the tortoise coordinate $r_{\ast}$,
we note that the potencial increases with decreasing temperature. However,
the shape of the potential does not change as one goes from high- to
intermediate-temperature values. Therefore, on basis of the behavior of $V$,
we do not expect a significative change in the spectrum of quasinormal modes
as a function of temperature, for all $\widetilde{T}^{2}\gtrsim 1$, in relation to
the $\Phi=0$ case.

\FIGURE{
\centering\epsfig{file=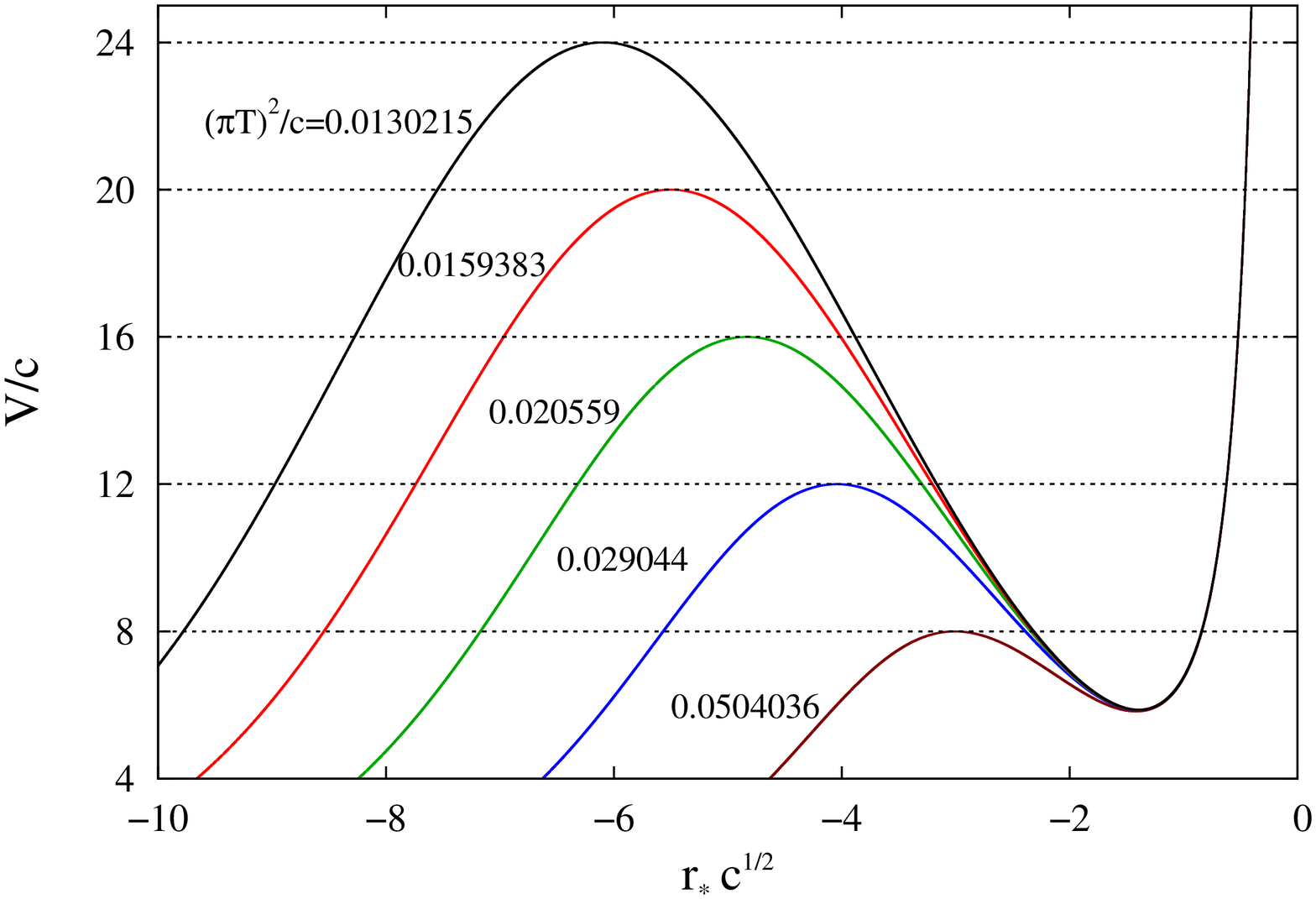,
width=7.2cm, angle=270}
\caption{The behavior of the potential for low temperature values and $q=0$. 
The horizontal dotted lines correspond to the masses of the glueball states at zero temperature.}
\label{potlowtemperature}}

\subsubsection*{Low Temperatures}

At very low temperatures ($\widetilde{T}^{2}\ll 1$), the effective
potential begins to develop a well near the AdS boundary. This is
illustrated in figure \ref{potlowtemperature} where we plot the potential
for some selected values of temperature and $q=0$. As the temperature
decreases the potential well gets deeper until it reaches the
zero-temperature form. In the regime of low temperatures, the
potential \eqref{potential} with vanishing wavenumber can be approximated by
\begin{equation}
V=\frac{1}{z^2}\left[\frac{15}{4}-\frac{3}{2}\frac{z^4}{z_{h}^{4}}+
2cz^{2}+c^{2}z^{4}-2c^{2}\frac{z^8}{z_{h}^{4}}+
{\cal{O}}\left(\frac{z^8}{z_h^8}\right)\right],
\label{Vreduzido2}
\end{equation}
where we have considered $(z/z_{h})^4$ to be small, keeping only first-order
terms in this parameter. The foregoing expression can be written in terms
of the temperature as
\begin{equation}
V=\frac{15}{4z^2}+2c+c^{2}z^{2}-\left(\frac{3}{2}+2c^{2}z^{2}\right)z^{6}\pi^{4}T^{4}\, ,
\label{Vreduzido3}
\end{equation}
where one can see explicitly the zero-temperature potential
\eqref{potentialzerotemp} plus low-temperature corrections, 
near the boundary. It is interesting to represent this potential
in terms of the tortoise coordinate $r_\ast$. Near the boundary
and for low temperatures, one can approximate the function
$z(r_{\ast})$ by
\begin{equation}
z=-r_{\ast}\left[1-\frac{r_{\ast}^{4}}{5z_{h}^{4}}\right].
\end{equation}
Then, the potential \eqref{Vreduzido3} reads
\begin{equation}
V=2c+c^{2}r_{\ast}^{2}+\frac{15}{4r_{\ast}^{2}}-
\frac{12}{5}c^{2}\,\pi^{4}\,r_{\ast}^{6}\,T^{4}\, .
\label{Vreduzido4}
\end{equation}
The above potential is a combination of a constant term $2c$, an oscilator-like
contribution proportional to $r_\ast^{2}$, an infinity barrier at $r_{\ast}=0$,
given by the term $15/4r_{\ast}^{2}$, and temperature dependent corrections.  
The negative sign of the thermal corrections, in the low-temperature regime, 
imply that the real part of the QN frequency $\omega$ will decrease with the
temperature. Indeed, we will observe this behaviour in the study of the quasinormal
modes in section \ref{spectrum}.

An important consequence of the presence of the potential well at very low
temperatures is the appearence of trapped modes. These modes are associated
with scalar glueballs and will show up as peaks in the spectral function
(the imaginary part of the Green's function) as will be discussed in the next
section.


\section{The retarded Green's function}
\label{Greenfunction}

In order to calculate the glueball retarded Green's function we follow the
prescription of Ref. \cite{Son:2002sd}. First we use the equation of motion
\eqref{eqwave1} to write down the on-shell version of the scalar field
action \eqref{action0}: 
\begin{equation}
S_{on\,shell}=\frac{\pi^{3} L^{5}}{4\kappa_{10}^2}\int d^{4}x\sqrt{-g}
e^{-\Phi}\,g^{zz}\,\phi\,\partial_{z}\phi\,{\vert}_{_{z=0}} \, ,
\label{action1}
\end{equation}
where we have disregarded the boundary term at the horizon $z=z_{h}$.
The Fourier transformed on-shell scalar field $\bar{\phi}(z,k)$ is
decomposed as 
\begin{equation}
\bar{\phi}(z,k)=\phi_k(z)\phi_0(k),
\label{decomp}
\end{equation}
where $\phi_k(z)$ is the  ``bulk to boundary propagator'' subjected to
the boundary condition
\begin{equation}
\lim_{z\to0}\phi_{k}(z)=1 \,.
\label{boundarycondition}
\end{equation}

According to the Minkowkskian prescription of Ref. \cite{Son:2002sd},
when computing the retarded Green's function one must consider a
propagator $\phi_k(z)$ which satisfy an incoming-wave condition at
horizon in the finite-temperature model,
\begin{equation}
\lim_{z \to z_h} e^{-\frac{cz^2}{2}} \left(\frac{z}{L}\right)^{-3/2}\phi_k(z)=
\frac{1}{\phi_{0}(k)}\,e^{-i\omega r_{\ast}} \,,
\label{horizoncondition}
\end{equation}
or a regularity condition\footnote{The zero temperature condition \eqref{regcondpsi}
is equivalent to imposing the regularity condition $\lim_{z\to\infty}\phi_k(z)=0$,
for space-like momentum ($k^2>0$), and then performing a Wick rotation, similar
to what was considered in Ref. \cite{Son:2002sd}.} at infinity ($z\to\infty$)
in the zero-temperature case,
\begin{equation}
\lim_{z \to \infty} e^{-\frac{cz^2}{2}} \left(\frac{z}{L}\right)^{-3/2}
\phi_k(z)=0\,.
\label{regcondpsi}
\end{equation}

\noindent Using the decomposition (\ref{decomp}), the action (\ref{action1}) 
takes the form
\begin{equation}
S_{on\,shell}=-\int\frac{d^{4}k}{(2\pi)^4}\,\phi_0(-k){\cal F}(k,z)
\,\phi_0(k)\,{\vert}_{_{z=0}} \, .
\label{action2}
\end{equation}
The retarded Green's function is then given by 
\begin{equation}
G^R(k)\equiv -2 \,{\cal F}(k,z)\,{\vert}_{_{z=0}}
=\frac{\pi^3 L^5}{2\kappa_{10}^2}  \lim_{z \to 0} \sqrt{-g}\,
g^{zz} e^{-\Phi} \phi^{\ast}_{k}(z) \partial_z \phi_k(z)\, .
\end{equation}

Since $\phi_k(z)$ is a solution of Eq. \eqref{eqwave2} satisfying the
normalization condition \eqref{boundarycondition} and additionally
the boundary condition \eqref{horizoncondition} in the
finite-temperature case and \eqref{regcondpsi} in the zero-temperature
case, it can expressed in terms of the wave functions defined in the
last section as 
\begin{equation}
\phi_k(z)=z^{3/2}e^{\frac{cz^2}{2}}\frac{\psi_{a}}{{\cal A}_{\ss{(a)}}} 
=z^{3/2}e^{\frac{cz^2}{2}}\left[\psi_{2}(z)+
\frac{{\cal B}_{\ss{(a)}}}{{\cal A}_{\ss{(a)}}}\psi_{1}(z)\right],
\end{equation}
where the index $a$ takes the value $0$ for the zero temperature
case and $ - $ for finite temperature. Using the expansions
\eqref{normalizable} and \eqref{nonnormalizable} for
$\psi_1$ and $\psi_2$, we find 
\begin{equation}
G^{R}(k)=\frac{N_{c}^{2}}{8\pi^2}\left[(c+2a_{21})\epsilon^{-2}+
4b\,\mbox{ln}(c\epsilon^2)
+2\left(b+a_{21}c+a_{21}^{2}\right)+4 {\rm Re}
{\frac{{\cal B}_{\ss{(a)}}}{{\cal A}_{\ss{(a)}}}}
-i\,{\rm Im}{\frac{{\cal B}_{\ss{(a)}}}{{\cal A}_{\ss{(a)}}}}\right],
\label{greenfunction}
\end{equation}
where $\epsilon$ is an ultraviolet regulator. In the above equation
we have used the relation 
\begin{equation}
\frac{\pi^3 L^8}{2 \kappa_{10}^2}=\frac{N_c^2}{8 \pi^2} \, . 
\end{equation}

The first two terms in Eq. \eqref{greenfunction} are ultraviolet divergent
and can be removed using a holographic renormalization procedure
like that of Ref. \cite{Bianchi:2001kw}. It is interesting to observe
that the imaginary part of the retarded Green's function at zero (finite)
temperature depends only on the quocient ${\cal B}_{\ss{(0)}}/{\cal A}_{\ss{(0)}}$
(${\cal B}_{\ss{(-)}}/{\cal A}_{\ss{(-)}}$). This quantity defines the
well-known spectral function which has been calculated in Ref.
\cite{Fujita:2009wc} for vector mesons in the soft-wall model,
and for other fields and holographic models, for example,
in Refs. \cite{Myers:2007we,Kovtun:2006pf, Teaney:2006nc,Myers:2008cj,Myers:2008me}.

\subsection{The zero-temperature case}

In the zero temperature case, the bulk to boundary propagator
has an analytical solution consisting on a combination of Kummer
and Tricomi confluent hypergeometric functions. The Kummer function
is ruled out by the regularity condition leading to 
\begin{equation}
\phi_k(z)=\Gamma \left(\frac{k^2}{4c} +2\right) \, c^2\, z^4
\, {\cal U} \left(\frac{k^2}{4c}+2 \, , 3 \, , \,cz^2 \right),
\label{zero-temp}
\end{equation}
where $k^{2}=-\omega^2+q^2$ and ${\cal U}(a,b,w)$ is the Tricomi function.
From the asymptotic expansion of the Tricomi function we find near the boundary:
\beqa
e^{-\frac{cz^2}{2}}\,z^{-\frac32}\,\phi_k(z)&\,=\,& z^{-\frac32}\,
\Big \{ 1 \,+\, a_{21} z^2 \,+\, b \,  z^4 \,\ln (c z^2) \nonumber \\
&\,+ \,& \Big[ \frac18 + \frac{k^2}{8c} - \frac{k^2}{8c}\,\Big(\frac{k^2}{4c}+1\Big)
\Big(\Psi\Big[\frac{k^2}{4c}+2\Big] - \Psi(1) - \Psi(3)\Big) \Big ]\,c^2\,z^4\nonumber\\
&\,+\,&  {\cal O} (z^6) \,+\, {\cal O}(\,z^6 \ln c z^2\, ) \Big \},
\eeqa
where $\Psi(x)$ is the digamma function and  $a_{21}$ and $b$ are given
in Eq. \eqref{coeffsol}. From the above expansion we can extract 
\beqa
\frac{{\cal B}_{\ss{(0)}}}{{\cal A}_{\ss{(0)}}}\,=\, \Big\{\frac18 + \frac{k^2}{8c}
-\frac{k^2}{8c}\, \Big(\frac{k^2}{4c}+1\Big) \Big[ \Psi\Big(\frac{k^2}{4c}+2\Big)
-\Psi(1) - \Psi(3) \Big] \Big \}\,c^2 \,.
\eeqa
Since the digamma function $\Psi(x)$ has poles at $x\,=\,-n$, it is possible
to extract an imaginary part of the retarded Green's function:
\beqa
{\rm Im }\, G^R_{_{T=0}}(k)&\,=\,&-\frac{N_c^2}{8 \pi^2}\,
{\rm Im}{\frac{{\cal B}_{\ss{(0)}}}{{\cal A}_{\ss{(0)}}}}\,=\,
\frac{N_c^2}{8 \pi^2}\,\frac{k^2}{8c}\,\Big(\frac{k^2}{4c}+1\Big)
\, \lim_{ \epsilon \to 0} \, {\rm Im} \Psi\Big(\frac{k^2}{4c}+2-i
\epsilon\Big)\nonumber\\
&\,=\,& \frac{N_c^2}{8 \pi}\,\frac{k^2}{8c}\,\Big(\frac{k^2}{4c}+1\Big)^2
\sum_{n=0}^{\infty} \frac{1}{n+1} \delta\Big( \frac{k^2}{4c} +2 + n\Big) \, . 
\label{imaggreenfuncion}
\eeqa
Hence, the spectral function at zero temperature is a sum of delta
functions localized at $k^2\,=\,- 4 c (n+2)$. These poles are interpreted
as the glueball masses at zero temperature. This result is consistent with
the Feynman correlator obtained in Ref. \cite{Colangelo:2007if}.

\subsection{Spectral function at finite temperature}
\label{finite-temp}

\subsubsection{General procedure}

In order to find the spectral function at finite temperature we must calculate the 
imaginary part of the retarded Green's function. We see from Eq.
\eqref{greenfunction} that $\mbox{Im}\,G^{R}(k)$ depends on the
ratio ${\cal B}_{\ss{(-)}}/{\cal A}_{\ss{(-)}}$, a quantity that can be
calculated by a numerical procedure consisting in the steps described below.

Firstly, we solve numerically the Schr\"odinger equation \eqref{Schrod1}
for a wave function that tends asymptotically to the normalizable solution
$\psi_{1}(z)$ as $z\to 0$. In other words, we integrate numerically this
equation from a point $z=z_{i}$ near the AdS boundary to a point $z=z_{f}$
near the horizon, using as ``initial'' conditions the truncated expansions
\begin{equation}
\psi_{1}(z_{i})=z_{i}^{5/2}\left(1+a_{11}z_{i}^{2}\right),\qquad
\partial_{z}\psi_{1}(z_{i})=\frac{1}{2}\,z_{i}^{3/2}\left(5+
9\,a_{11}z_{i}^{2}\right).
\label{initialcond1}
\end{equation}
The same numerical integration procedure is employed
for a wave function that tends to the non-normalizable solution
$\psi_2(z)$ as $z\to 0$. In this case, we use the following
conditions near the boundary:
\begin{equation}
\begin{gathered}
\psi_{2}(z_{i})=z_{i}^{-3/2}\left(1+a_{21}z_{i}^{2}\right)+
b\,\psi_{1}(z_{i})\,\ln{(c z_{i}^{2})},\\
\partial_{z}\psi_{2}(z_{i})=\frac{1}{2}\,z_{i}^{-5/2}\left[-3+a_{21} z_{i}^{2}\right]
+b\,\partial_{z}\psi_{1}(z_{i})\,\ln{(c z_{i}^{2})}+\frac{2b}{z_{i}}\,\psi_{1}(z_{i}).
\end{gathered}
\label{initialcond2}
\end{equation}
The expressions \eqref{initialcond1} and \eqref{initialcond2} are obtained
from the near boundary asymptotic behavior for $\psi_{1}$ and $\psi_{2}$
discussed in the last section.

With the numerical results for $\psi_{1}$ and $\psi_{2}$ at
$z=z_{f}$ we construct the vectors 
\beqa
\begin{pmatrix}
\psi_{j}(z_f) \\
\partial_{z}\psi_{j}(z_f) 
\end{pmatrix} \,=\, \begin{pmatrix}
\psi_-(z_f) & \psi_+(z_f) \\
\partial_{z}\psi_{-}(z_f) & \partial_{z}\psi_{+}(z_f)
\end{pmatrix} \begin{pmatrix} {\cal C}_{\ss{(j)}}
\\ {\cal D}_{\ss{(j)}}
\end{pmatrix}\qquad\qquad
(j=1,2)\, ,
\label{nearhorizoncoeff} 
\eeqa
with $\psi_{\pm}(z_f)$ and $\partial_{z}\psi_{\pm}(z_f)$ given
by the near-horizon expansions \eqref{nearhorizonexp}
truncated at second order in $(1-z_{f}/z_{h})$.
Then we extract the coefficients ${\cal C}_{\ss{(j)}}$ and
${\cal D}_{\ss{(j)}}$ by inverting \eqref{nearhorizoncoeff}:
\beqa
\begin{pmatrix} {\cal C}_{\ss{(j)}} \\ {\cal D}_{\ss{(j)}}\end{pmatrix}
 \,=\, \begin{pmatrix} \psi_-(z_f) & \psi_+(z_f) \\
\partial_{z}\psi_-(z_f) & \partial_{z}\psi_+(z_f)
\end{pmatrix}^{-1} \begin{pmatrix} \psi_{\ss{(j)}}(z_f) \\
\partial_{z}\psi_{\ss{(j)}}(z_f) 
\end{pmatrix}
\qquad\qquad (j=1,2).
\label{coeficients2}
\eeqa
Finally, from Eqs. \eqref{coeffrelation} and \eqref{coeficients2}
we obtain 
\beq
\frac{{\cal B}_{\ss{(-)}}}{{\cal A}_{\ss{(-)}}}=
-\frac{{\cal D}_{\ss{(2)}}}{{\cal D}_{\ss{(1)}}}
=\frac{\partial_{z}\psi_{-}(z_f)\psi_2(z_f)-
\psi_{-}(z_f)\partial_{z}\psi_{2}(z_f)}{\partial_{z}\psi_{-}(z_f)
\psi_{1}(z_f)-\psi_{-}(z_f)\partial_{z}\psi_1(z_f)}.
\eeq
This way we can obtain numerically  ${\cal B}_{\ss{(-)}}/{\cal A}_{\ss{(-)}}$
using the numerical results for $\psi_1$, $\psi_2$, $\partial_{z}\psi_1$
and $\partial_{z}\psi_2$ at a point $z=z_f$ near the horizon. 

\subsubsection{Numerical results}

The imaginary part of the retarded Green's function
gives us the spectral function
\beq
{\cal R}(\omega,q) \,\equiv\, - 2  \, {\rm Im }\, G^R(\omega, q)
\,=\,\frac{N_c^2}{4 \pi^2}\,{\rm Im}
\frac{{\cal B}_{\ss{(-)}}}{{\cal A}_{\ss{(-)}}} \,. 
\eeq
\noindent
We present here some numerical results for the spectral function
obtained using the procedure discussed above. We show in figures
\ref{spectralfunction1} and \ref{spectralfunction2} the dependence
of the spectral function with the frequency $\widetilde{\omega}=\omega/\sqrt{c}$
for selected values of temperature $\widetilde{T}=\pi T/\sqrt{c}$
and momentum $\widetilde{q}=q/\sqrt{c}$. We can see the presence of various peaks
that correspond to the poles of the retarded Green's function. As commented above,
these poles are associated with the frequencies of the black-hole quasinormal
modes and are interpreted as glueball states in the dual gauge theory.
Indeed, near one of the peaks the spectral function can be approximated
by a Lorentzian function:
\beq
{\cal R}(\widetilde{\omega},\widetilde{q}) \sim
\frac{\widetilde{\omega}_{\ss{I}}}{(\widetilde{\omega}
-\widetilde{\omega}_{\ss{R}})^2 +\widetilde{\omega}_{\ss{I}}^2}\,,
\label{lorentzian}
\eeq
where $\widetilde{\omega}_{\ss{R}}$ and $\widetilde{\omega}_{\ss{I}}$ are
respectively the real part and the negative of the imaginary part of
the black-hole quasinormal frequencies. The half-width of the distribution
\eqref{lorentzian}, given by $\widetilde{\omega}_{\ss{I}}$, is interpreted in the dual
gauge theory as the inverse of the glueball lifetime (multiplied by the constant 
$\sqrt{c}\,$). The real part $\widetilde{\omega}_{\ss{R}}$, for $q=0$, gives a measure
of the glueball mass at finite temperature.

\FIGURE{
\centering\epsfig{file=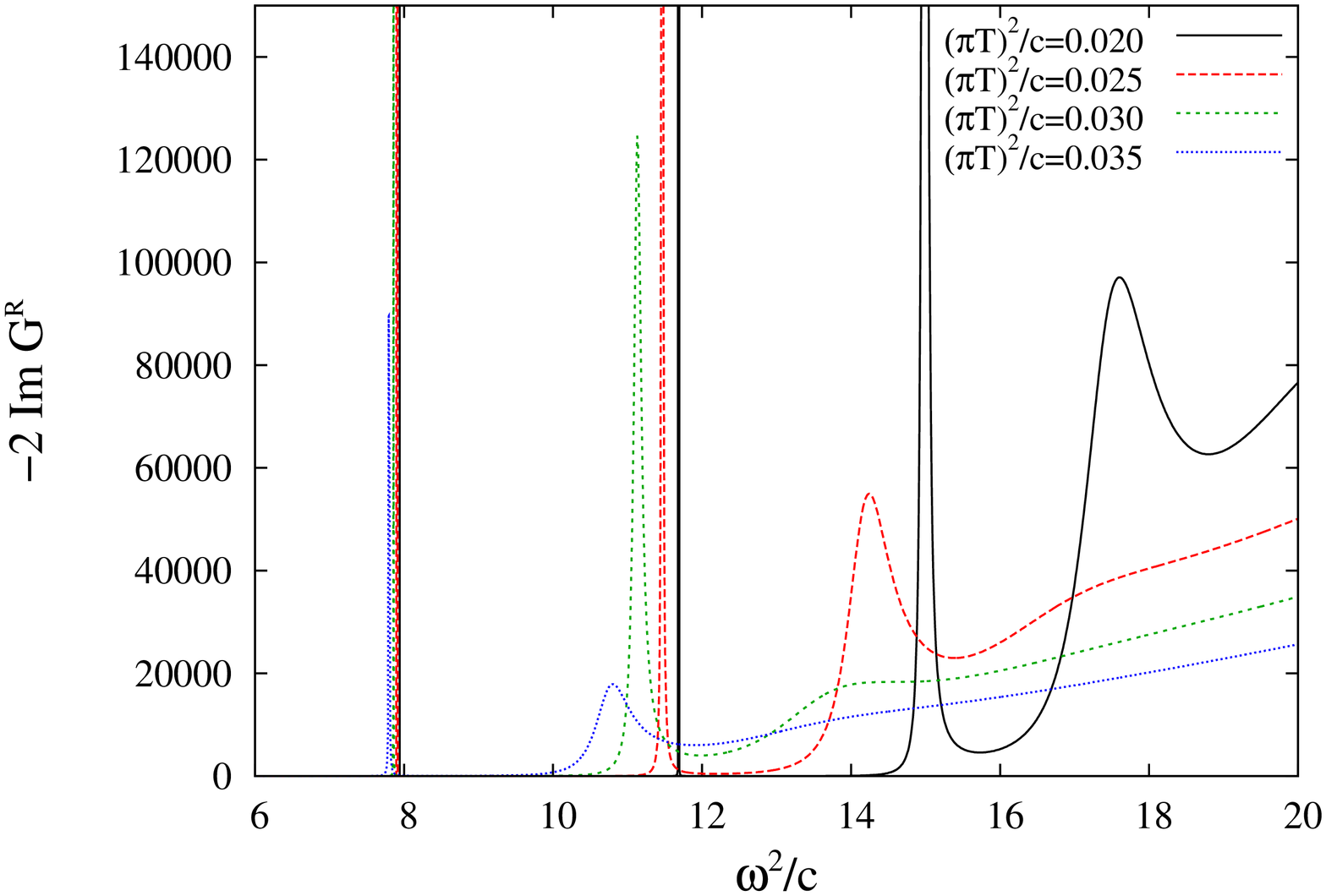,
width=8.15cm, angle=270}
\caption{Spectral functions for $q=0$ and selected values of temperature in
units of $N_{c}^{2}/4\pi^{2}$. Note that the first peak of the spectral
functions for the four cases almost coincide around $\widetilde{\omega}^2=8$.}
\label{spectralfunction1}}

We consider in Fig. \ref{spectralfunction1} the zero-momentum case for various
values of temperature. Note that as the temperature increases the number of peaks
in the spectral function decreases while their width increase. Accordingly, the
number of glueball quasiparticle states and their lifetimes decrease
with the temperature. If we continue increasing the temperature we will reach
a phase (for $\widetilde{T}^{2}\gtrsim 0.1$) in which there are no more peaks.
This suggests a kind of {\it glueball melting} in the sense that the glueball
excitations disappear. A similar result was recently obtained in Ref. \cite{Fujita:2009wc} 
when studying finite-temperature effects on the spectrum of vector mesons in the
soft-wall model. It is important to remark that for the range of temperatures
in which we found glueball excitations (peaks in the spectral function),
the theory is in an unstable phase. 

It is interesting to analyze the spectral function of Fig.
\ref{spectralfunction1} on basis of the behavior of the effective
potential of Fig. \ref{potlowtemperature}. For instance, for
$\widetilde{T}^{2}=0.02$ there are three modes ``trapped'' in
the potential well, which are represented by dotted lines.
These quasi-stationary trapped modes correspond to the three sharp peaks
that appear in the spectral function ${\cal R}(\widetilde{\omega},0)$.
Analogously, for $\widetilde{T}^{2}=0.03$, the two sharp peaks at
$\widetilde{\omega}^2\approx 8$ and $\widetilde{\omega}^2\approx 11$
are associated to the modes trapped in the potential well.

Note that the dotted lines shown in the effective potential graph of 
Fig. \ref{potlowtemperature} correspond to the zero temperature modes. 
However, as shown in Fig. \ref{spectralfunction1}, the masses of the glueballs, 
given by the positions of the peaks, vary with the temperature. Nevertheless, this
effect does not change the qualitative analysis made above.

Another important feature shown in Fig. \ref{spectralfunction1} is that, 
for a fixed temperature, the width of the peaks increases with the frequency. 
This can be explained  in terms of the tunneling effect of the scalar-field
waves in the potential barriers of Fig. \ref{potlowtemperature},
which is enhanced when the frequency becomes comparable with the
height of the potential well. Then, the glueball states at finite
temperature become less stable as the energy increases. 

\FIGURE{
\centering\epsfig{file=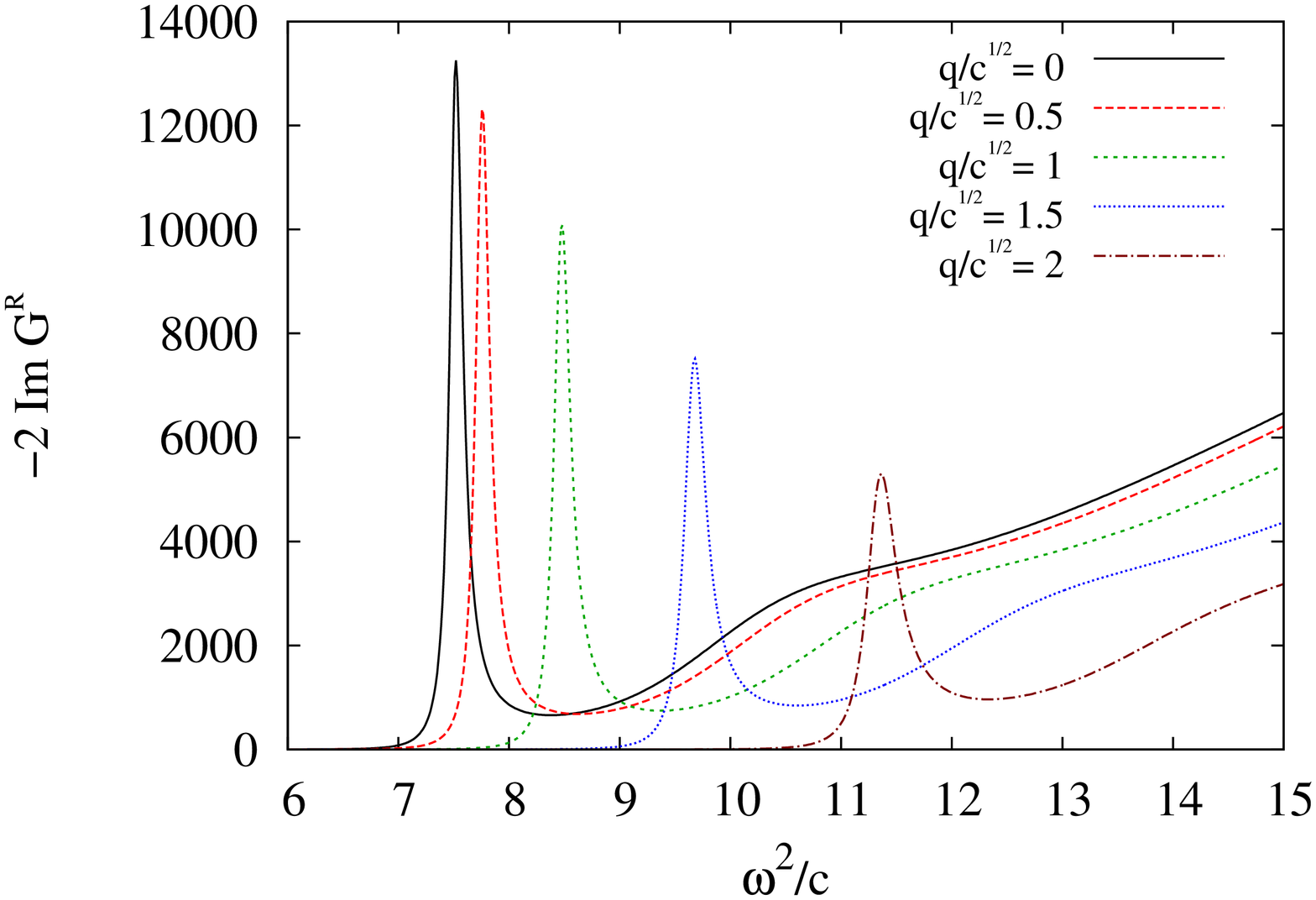,
width=8.15cm, angle=270}
\caption{Spectral function for $\widetilde{T}^{2}=0.05$ and selected values
of momentum $\widetilde{q}$ in units of $N_{c}^{2}/4\pi^{2}$.}
\label{spectralfunction2}}

In figure \ref{spectralfunction2}, we plot the spectral function for a
fixed value of temperature $\widetilde{T}^2=0.05$ and various values of
wavenumber $\widetilde{q}$. In this picture one can see that the position
and the width of the peak increases with the momentum.
This result indicates that the glueball energy increases with the momentum,
as expected, while the glueball lifetime decreases.

In this section the frequency $\omega$ was taken as a real quantity for
the computation of the spectral function ${\cal R}(\widetilde{\omega},\widetilde{q})$.
However, when studying the black-hole quasinormal modes (the poles of $G^{R}$)
in the next section, we will look for frequencies of the form
$\widetilde{\omega}_{\ss{QN}}=\widetilde{\omega}_{\ss{R}}
-i\,\widetilde{\omega}_{\ss{I}}$ with $\widetilde{\omega}_{\ss{R}}$ and
$\widetilde{\omega}_{\ss{I}}$ being (positive) real quantities.


\section{Quasinormal modes and the scalar glueballs at finite $T$}
\label{spectrum}

Quasinormal modes (QNMs) are solutions for classical field perturbations on a
black hole geometry (see Refs. \cite{Berti:2009kk,Kokkotas:1999bd,Nollert:1999ji}
for general reviews on the subject). These solutions are subjected to an absorption
condition at the horizon and a regularity condition at the spacetime boundary
in the AdS black hole case. The absorption condition leads to a purely
ingoing wave at the horizon while the regularity at the boundary translates 
into a Dirichlet condition for $\psi$, for the case of a massless scalar field discussed here. 

The analysis of the effective potential in section \ref{model} gives additional
support to the above quoted boundary conditions. The fact that the effective
potential diverges at the spacetime boundary $z=0$ makes it natural to impose 
a Dirichlet condition, while the vanishing of the potential at $z=z_{h}$
and the one-way membrane property of the horizon lead naturally to ingoing
plane wave solutions. 

The absorption condition can be implemented by means of the solution
$\psi_{\ss{-}}(z)$ of section \ref{model}, which behaves like
an ingoing wave at the horizon, according to Eq. \eqref{nearhorizonexp}.
Imposing Dirichlet boundary condition on $\psi_{\ss{-}}(z)$ at $z=0$ implies that the
connection coefficient $\mathcal{A}_{\ss{(-)}}(\omega,q)$ vanishes.
Since the retarded Green's function \eqref{greenfunction} diverges when
$\mathcal{A}_{\ss{(-)}}(\omega,q)=0$, we conclude
that the {\it black-hole quasinormal modes} correspond to the (complex)
poles of the glueball retarded Green's function. 

In the following, we will show the numerical results obtained for the complex
roots $\omega_{\ss{QN}}=\omega_{\ss{R}}-i\,\omega_{\ss{I}}$ of the equation
$\mathcal{A}_{\ss{(-)}}(\omega,q)=0$. These complex roots are the frequencies
of the scalar-field quasinormal modes of the black hole in the presence
of the soft-wall dilaton background. Previous studies of quasinormal modes
in the context of gauge/gravity duality can be found, for example, in Refs.
\cite{Horowitz:1999jd,Wang:2000dt,Cardoso:2001hn,Cardoso:2001bb,
Starinets:2002br,Nunez:2003eq,Kovtun:2005ev,Maeda:2005cr,Siopsis:2004as,
Miranda:2005qx,Zhang:2006bc,Hoyos:2006gb,Miranda:2007bv,Amado:2008hw}.

\subsection{Methods}

In the last decades, different methods have been developed to numerically
calculate the quasinormal modes of black holes. We employ here two of
these techniques to find out the QNMs of a scalar field in the soft-wall model.
The first one is the traditional Fr\"obenius power series method, following the
approach of Horowitz and Hubeny \cite{Horowitz:1999jd}. In this case the problem of
finding quasinormal frequencies is reduced to that of finding the roots of
a polynomial equation. The second method is based on ideas from the study of
the ressonant scattering problem in atomic and nuclear
physics. It has been recently applied to the calculation of QNMs 
\cite{Berti:2009wx} and is presently known as the Breit-Wigner resonance method.

\subsubsection{The power series method}

The power series method of Ref. \cite{Horowitz:1999jd} is of particular
interest for the present work since it is suited for asymptotically
AdS spacetimes. We start with the scalar field equation of motion written
as  a Schr\"odinger-like equation \eqref{Schrod1}. As discussed above, QNMs are
defined as solutions which are purely ingoing waves at horizon,
$\psi\rightarrow {\rm e}^{-i\omega r_{\ast}}$, and satisfy a Dirichlet
condition at $z=0$. In the power series method, it is convenient
to introduce a new function $\varphi=e^{+i\omega r_{\ast}}\psi$. Then, 
Eq. \eqref{Schrod1} becomes
\begin{equation}
f(z)\,\frac{d^{\,2}\varphi}{dz^{2}} \,+\,\left(2i\omega-4\frac{z^{3}}{z_{h}^{4}}\right)
\,\frac{d\varphi}{dz} \,-\, \frac{V(z)}{f(z)}  \,\varphi=0\,.
\label{fund-z}
\end{equation}
The above equation has a regular singular point ($z=z_{h}$) in the region of
interest, $0\leq z\leq z_{h}$. So, according to the Fuchs's theorem,
at least one of the solutions of Eq. \eqref{fund-z} can be written as a
Fr\"obenius series \cite{arfken}. In fact, from the solution
\eqref{nearhorizonexp} for $\psi_{-}$ one can see that the corresponding
function $\varphi_{-}$ can be written as a power series of the form:
\begin{equation}
\varphi_{\ss{-}}(z)=\sum_{m=0}^{\infty}a_{m{\ss(-)}}\left(1-\frac{z}{z_{h}}\right)^{m}.
\label{serie}
\end{equation}
\noindent
Substituting Eq. \eqref{serie} into \eqref{fund-z} we obtain a recurrence
relation for the coefficients $a_{m}$'s. Since Eq.  \eqref{fund-z} is linear
we can set $a_0=1$ and calculate this way any coefficient $a_{m{\ss(-)}}$
as a function of the frequency $\widetilde{\omega}=\omega/\sqrt{c}$,
the wavenumber $\widetilde{q}=q/\sqrt{c}$, and the temperature
$\widetilde{T}=\pi T/\sqrt{c}$. The coefficients $a_{1{\ss(-)}}$
and $a_{2{\ss(-)}}$ are given explicitly in equation \eqref{coefa1a2}.
Then, imposing the Dirichlet condition at AdS boundary, we obtain
\begin{equation}
\varphi_{\ss{-}}(0)=\sum_{m=0}^{\infty}a_{m{\ss(-)}}=0.
\label{eq-serie}
\end{equation}
Thus, the problem of finding quasinormal (QN) frequencies has been
reduced to that of obtaining the roots of the equation
\eqref{eq-serie}. Since we cannot carry out numerically the full
sum in expression \eqref{eq-serie}, we truncate the series after
a large number of terms and look for the zeros of the resulting partial sum.
The precision of the results is then verified through the relative
variation between the roots of two successive partial sums. For each
pair of $\widetilde{q}$ and $\widetilde{T}$, we find a set
of QN frequencies $\widetilde{\omega}$, labeled by the principal
quantum number $n$ and ordered in a set beginning by the roots with
the lowest absolute values for the imaginary part.

\subsubsection{The Breit-Wigner method}

Although the power series method is suitable to find QN frequencies
in the intermediate- and high-temperature regimes ($\widetilde{T}\gtrsim 1$),
its convergence gets weak at very low temperatures ($\widetilde{T}\ll 1$).
Fortunately, in this regime the behavior of the effective potential 
$V(r_{\ast})$ enable us to use an alternative method based on the
Breit-Wigner theory of resonances. 

The theory of Breit-Wigner resonances was originally employed in
gravitational problems to find the resonant frequencies associated 
with ultra-compact relativistic stars \cite{Thorne:1969,ChandrasekharFerrari,
Chandrasekhar:1992ey}, and only recently this technique was used to find
QN frequencies of black holes, particularly of very small Schwarzschild-AdS
black holes \cite{Berti:2009wx,Berti:2009kk}. An important element for the
efficiency of this method is the form of the effective potential:
$V(r_{\ast})$ should have the form of a very deep well, so that 
quasi-stationary ``trapped'' states can exist and the scalar field
waves can only leak out to the horizon by tunneling through the potential
barrier. This is exactly the situation found in section \ref{model}
when analyzing the effective potential for a massless scalar field
in the soft-wall black hole geometry. Below we present the general
idea of this method in the form used in asymptoticaly AdS spacetimes.

The resonance method deals with normalizable solutions, which
corresponds here to choosing the solution $\psi_1$ of Eq. \eqref{normalizable}.
As discussed in section \ref{model}, the function $\psi_{1}$ can be written
as a combination of {\textit{ingoing}} and {\textit{outgoing}} waves $\psi_{\pm}$,
as shown in Eq. \eqref{inverserelation}. This equation determines
the asymptotic behavior of $\psi_{1}$ at the horizon:
\begin{equation}
\psi_{1}=\mathcal{C}_{1}e^{-i\omega r_{\ast}}+
\mathcal{D}_{1}e^{i\omega r_{\ast}}=
\alpha\cos\omega r_{\ast}+\beta\sin\omega r_{\ast},
\end{equation}
where $\alpha=\mathcal{C}_{1}+\mathcal{D}_{1}$ and
$\beta=-i(\mathcal{C}_{1}-\mathcal{D}_{1})$. 
The QNMs are normalizable solutions with complex frequencies
$\omega_{\ss{QN}}=\omega_{\ss{R}}-i\,\omega_{\ss{I}}$ such that
there are no outgoing waves. This corresponds to set $\mathcal{D}_{1}=0$. 
Analogously, the complex conjugate of a QNM corresponds to the 
vanishing of the ingoing coefficient ($\mathcal{C}_{1}=0$). 
Assuming that the QNMs are simple zeros of $\mathcal{D}_{1}$
we expect that near these modes $\mathcal{D}_{1} \sim (\omega -\omega_{\ss R})
+i\omega_{\ss I}$ and  $\mathcal{C}_{1} \sim (\omega -\omega_{\ss{R}})
-i\omega_{\ss{I}}$. Then it turns convenient to introduce the real quantity 
\begin{equation}
\alpha^{2}+\beta^{2}=4\mathcal{C}_{1}\mathcal{D}_{1}\,\approx\,\mbox{const.}
\times\left[(\omega-\omega_{\ss{R}})^{2}+\omega_{\ss{I}}^{2}\right].
\label{waveasymptotic}
\end{equation}
The Breit-Wigner resonance method is based on the minimization of this
quantity, considering $\omega$ as a real variable
\cite{Thorne:1969,Chandrasekhar:1992ey}. Note that this is far simpler
than solving the complex equation $\mathcal{D}_{1}=0$ when using
the numerical integration procedure of section \ref{Greenfunction}.

The quantity $\alpha^{2}+\beta^{2}$ will present sharp dips at
$\omega=\omega_{\ss{R}}$ corresponding to the real parts of the
QN frequencies. Once we have located a minimum of $\alpha^{2}+\beta^{2}$
on the line of real $\omega$ at some $\omega=\omega_{\ss{R}}$, the imaginary
part of the QN frequency $\omega_{\ss I}$ is obtained
by means of a parabolic fit of the Breit-Wigner formula \eqref{waveasymptotic}.
To set the accuracy of our numerical results, we
use the alternative (and equivalent) expressions
for $\omega_{\ss I}$:
\begin{equation}
\omega_{\ss{I}}=-\beta/\alpha^{\prime}=\alpha/\beta^{\prime}, 
\label{betaalpha}
\end{equation}
where a prime indicates a derivative with respect to $\omega_{\ss{R}}$,
evaluated at the minimum \cite{ChandrasekharFerrari,Berti:2009wx}.
The Breit-Wigner method has, in general, a good convergence
as long as the condition $\omega_{\ss{I}} \ll \omega_{\ss{R}}$ is satisfied.

It is interesting to notice the relation between the dips of
$\alpha^{2}+\beta^{2}$ and the peaks of the spectral function
${\cal R}(\omega,q)$, discussed in section \ref{Greenfunction}.
Each minimum of \eqref{waveasymptotic} appears as a sharp
maximum of the spectral function. This connection can be
understood by considering the approximate expression
\beq
{\cal R}=-2\mbox{Im}G^{R}\sim (\mathcal{A}_{\ss{(-)}})^{-1} \sim(\mathcal{D}_{1})^{-1}\,.
\eeq
Therefore, the simple pole hypothesis for the retarded Green's function
that leads to \eqref{lorentzian} is directly related to the simple
zero hypothesis for the coefficient $\mathcal{D}_{1}$.

\subsection{Numerical results for the QN frequencies}

\subsubsection{Temperature dependence}

Here we present our numerical results for the QN frequencies when varying
the temperature while keeping the momentum $q=0$. The power series method
can be applied to a wide range of values from high temperatures
to $\widetilde{T}^2 \approx 0.022$. At lower temperatures, the power series
method has convergence problems but, as we discussed above, it is precisely
in this regime that the wells of the potential are deep and the dips in 
$\alpha^2+\beta^2$ are sharp enough to make the Breit-Wigner
method efficient. We have used the Breit-Wigner method for
$\widetilde{T}^{2}\lesssim 0.05$. A comparison of the results obtained with
the two methods is presented in Table \ref{tabcomparacao}. The approaches
are complementary, and show a very good agreement for the range
$0.025 < \widetilde{T}^2 < 0.04\,$. 

\TABLE{
\begin{tabular}{ccccc}
\hline\hline
& \multicolumn{2}{c}{Power series} & \multicolumn{2}{c}
{Breit-Wigner}\\
\cline{2-5}
\vspace{-0.45cm}\\
$\;\;\;\widetilde{T}^{2}\;\;\;$ & $\quad\;\;\;\widetilde{\omega}_{\ss{R}}^{2}
\quad\;\;\;$ & $\quad\;\;\;\widetilde{\omega}_{\ss{I}}^{2}\quad\;\;\;$ &
$\quad\;\;\;\widetilde{\omega}_{\ss{R}}^{2}\quad\;\;\;$ &
$\quad\;\;\;\widetilde{\omega}_{\ss{I}}^{2}\quad\;\;\;$\\
\hline
0.0493827 & 7.52779 & $1.41127\times 10^{-4}$ & 7.53270 & $1.43646\times 10^{-4}$ \\
0.04 & 7.72086 & $1.31516\times 10^{-6}$ & 7.72092 & $1.31549\times 10^{-6}$\\
0.0330579 & 7.82420 & $2.12446\times 10^{-9}$ & 7.82420 & $2.12446\times 10^{-9}$\\
0.0277778 & 7.88073 & $8.61149\times 10^{-13}$ & 7.88073 & $8.61151\times 10^{-13}$\\
0.0236686 & 7.91529 & $1.04760\times 10^{-16}$ & 7.91529 & $1.04873\times 10^{-16}$\\
0.0204082 & --- & --- & 7.93787 & $2.23960\times10^{-20}$\\
\hline\hline
\end{tabular}
\centering
\caption{A comparison between the values obtained for the
fundamental ($n=0$) QN frequency computed using the power
series and the Breit-Wigner resonance methods.}
\label{tabcomparacao}}

The values of $\widetilde{\omega}_{\ss{I}}$ shown in Table \ref{tabcomparacao}
for the Breit-Wigner method were calculated using the parabolic fit
of Eq. \eqref{waveasymptotic} for $\alpha^2+\beta^2$. The alternative
expressions \eqref{betaalpha} were used to check the consistency of
the results in the range of temperature values where the power series
method does not converge. In this case, the number of significant figures
retained in $\widetilde{\omega}_{\ss{I}}$ was chosen so that all the alternative
ways of determining $\widetilde{\omega}_{\ss{I}}$ give the same value.

\FIGURE{
\centering\epsfig{file=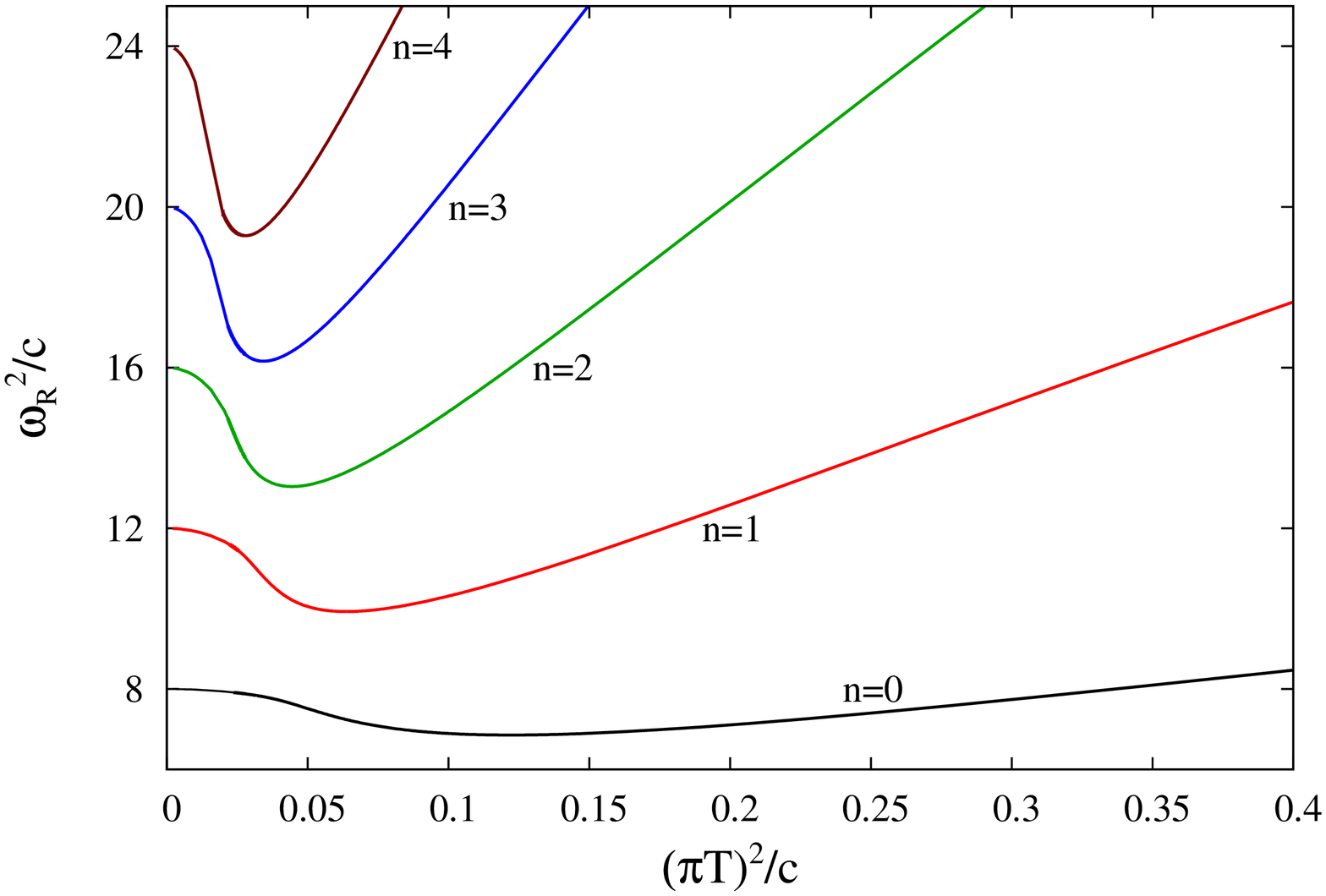,
width=5.15cm, angle=270}
\centering\epsfig{file=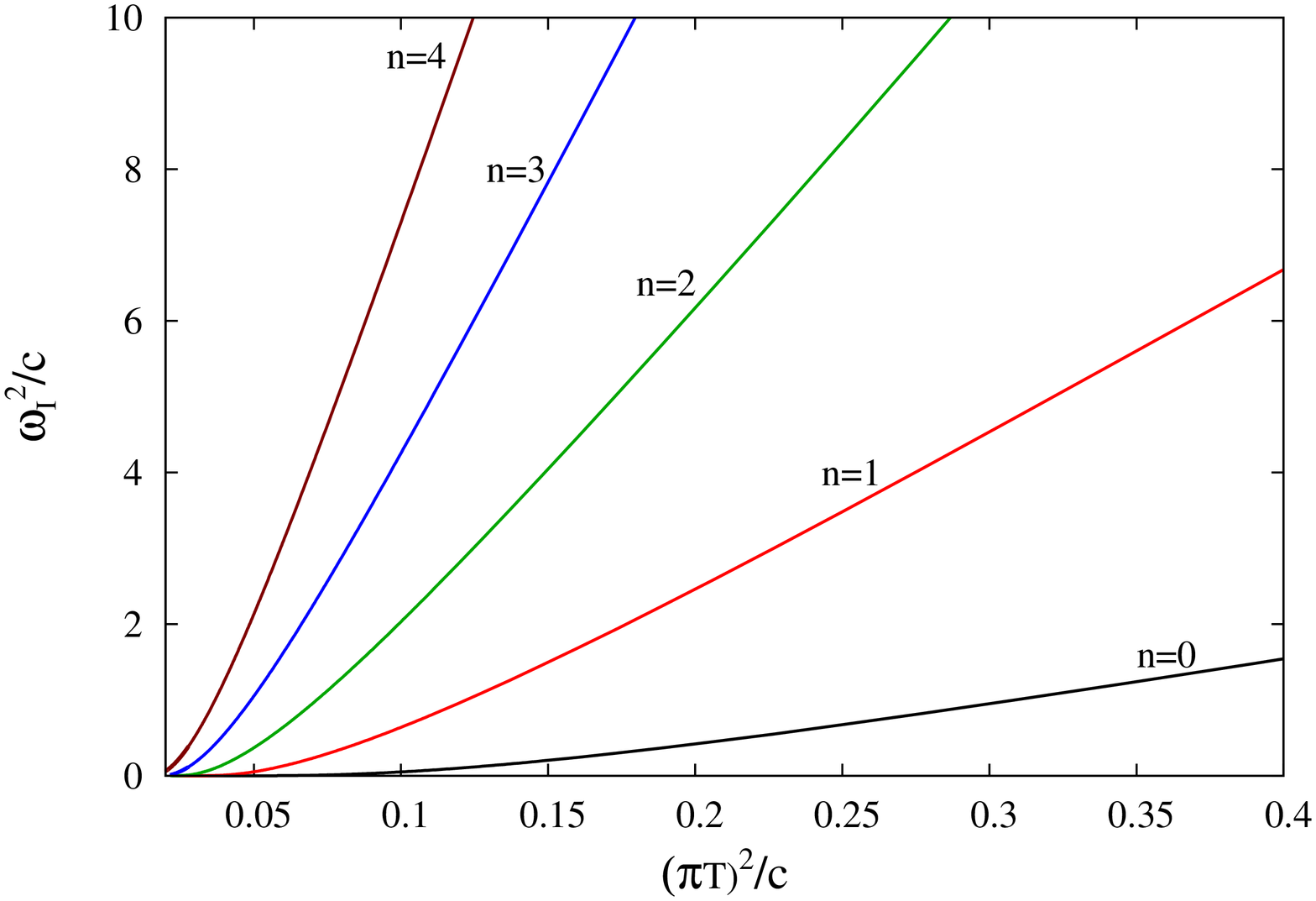,
width=5.15cm, angle=270}
\caption{Numerical results for the square of the real and imaginary
parts of the QN frequencies,  ${\omega}_{\ss{R}}^{2}/c$ and
${\omega}_{\ss{I}}^{2}/c$, for the first five quasinormal modes
$n=0,1,...,4$, with $q=0$.}
\label{QNfreqTemperature}}

In Fig. \ref{QNfreqTemperature}, we show our results for the square
of the real and imaginary parts of the QN frequencies as functions
of $\widetilde{T}^{2}$ for the first five modes $n=0,$ $1,\,...,\,4$
with $q=0$. We observe that, in the zero temperature limit,
the real part of the QN frequencies coincide with the corresponding
glueball mass spectrum, given by Eq. \eqref{T0spectrum}. Then, increasing
the temperature $\widetilde{\omega}_{\ss{R}}^{2}$ decrease until they reach
a minimum value $(\widetilde{\omega}_{\ss{R}}^{2})^{min}$ at some critical
temperature, different for each mode. The value of the critical temperature
is lower for higher overtone numbers $n$. For higher temperatures,
$\widetilde{\omega}_{\ss{R}}$ increases approaching a linear dependence
on the temperature. On the other hand, we observe in Fig.
\ref{QNfreqTemperature} that (the negative of) the imaginary part of
the frequencies $\widetilde{\omega}_{\ss{I}}$ increase monotonically
with the temperature approaching constant slopes. In Table \ref{tabminimum},
we show the values of the minimum of the curve $\widetilde{\omega}_{\ss{R}}^{2}$
versus $\widetilde{T}^{2}$ for the first five quasinormal modes. We also
show there the numerical values for $\widetilde{\omega}^2_{\ss{I}}$ and
the quality factor $Q=\omega_{\ss{R}}/2\omega_{\ss{I}}$ evaluated
at the critical temperatures. 
Notice that the values of $Q$ in the minimum of $\widetilde{\omega}_{\ss{R}}$ 
are such that $Q \lesssim 4\,$ and so, for temperatures equal or higher than 
the critical value, there are no more quasiparticle peaks corresponding to glueballs 
in the spectral function $\mathcal{R}(\widetilde{\omega},0)$. Quasiparticle states 
show up for higher values of $Q$.

\TABLE{
\begin{tabular}{ccccc}
\hline
\hline
\vspace{-0.45cm}\\
$\;\;\;n\;\;\;$ & $\quad\;\;\;\widetilde{T}^{2}\quad\;\;\;$ &
$\quad\;(\widetilde{\omega}_{\ss{R}}^{2})^{min}\quad\;$ &
$\quad\;\;\;\widetilde{\omega}_{\ss{I}}^{2}\quad\;\;\;$ &
$\quad\;\;\;Q\quad\;\;\;$\\
\hline 
0 & 0.122069  & 6.85489 & 0.108102 & 3.98156 \\
1 & 0.0637774 & 9.92861 & 0.166347 & 3.86284 \\
2 & 0.0444836 & 13.0406 & 0.241963 & 3.67066 \\
3 & 0.0343047 & 16.1605 & 0.319899 & 3.55378 \\
4 & 0.0279468 & 19.2844 & 0.398241 & 3.47936\\
\hline\hline
\end{tabular}
\caption{The minimum value of the real part of QN frequency
for the first five modes, $n=0,1,...,4$, and the corresponding values
of $\widetilde{\omega}_{\ss{I}}^{2}$, $\widetilde{T}^{2}$,
and the quality factor $Q$.}
\label{tabminimum}}

The decreasing of the real part of the quasinormal frequencies at very low
temperatures, shown in figure \ref{QNfreqTemperature}, is related to the fact
that the very low temperature potential given by \eqref{Vreduzido4} behaves
as a harmonic oscillator with an infinite barrier $r_\ast^{-2}$ and a
temperature dependent negative correction $\,-T^4r_\ast^6$. This
negative correction is responsible for the observed decrease in
$\widetilde{\omega}_{\ss{R}}^{2}$ for $\widetilde{T}^{2}\lesssim 0.1\,$.

The exact way the functions $\widetilde{\omega}_{\ss{R}}^{\,2}
(\widetilde{T}^{\,2})$ and $\widetilde{\omega}_{\ss{I}}^{\,2}
(\widetilde{T}^{\,2})$ deviate from the linear behavior, observed
in high temperatures, varies with the level $n$.
Within a certain accuracy, the frequency of each mode deviates significantly
from a straight line only in the low temperature regime. For instance, 
this deviation for the fundamental mode occurs below $\widetilde{T}^{2}=0.25\,$. 
For higher overtones, as $n$ increases, the deviations from the
linear behavior occur at decreasing temperatures.

We now compare our results for the soft-wall model with the pure
AdS black-hole case, which corresponds to the limit $\Phi=0$. 
Some results  for the fundamental mode ($n=0$) obtained with the Horowitz-Hubeny 
power series method are shown in Table \ref{tabtemperatura} for $q=0$ and selected 
values of temperature $\widetilde{T}$. 
We introduce for the real and imaginary parts of the QN frequencies 
the fractional differences
\begin{equation}
\Delta_{\ss{R,I}}=\frac{|\omega_{\ss{R,I}}-\omega_{\ss{R,I}}^{\ss{\Phi=0}}|}
{\omega_{\ss{R,I}}^{\ss{\Phi=0}}}=\frac{|\mathfrak{w}_{\ss{R,I}}
-\mathfrak{w}_{\ss{R,I}}^{\ss{\Phi=0}}|}
{\mathfrak{w}_{\ss{R,I}}^{\ss{\Phi=0}}}\,, 
\end{equation}
where $\mathfrak{w}=\omega/2\pi T$ is the normalized frequency
commonly used in the literature \cite{Nunez:2003eq,Kovtun:2005ev,
Kovtun:2006pf} and the superscript $\Phi=0$ indicates the results
for the AdS black hole without dilaton. Since the frequencies
$\mathfrak{w}_{\ss{R}}^{\ss{\Phi=0}}$ and
$\mathfrak{w}_{\ss{I}}^{\ss{\Phi=0}}$ are constants independent of
the temperature, for a fixed level $n$ the fractional differences
$\Delta_{\ss{R,I}}$ varies with $\widetilde{T}$ due to the changes in
the values of $\mathfrak{w}_{\ss{R,I}}=
\widetilde{\omega}_{\ss{R,I}}/2\,\widetilde{T}\,$.

As expected, one can see a significant difference in the numerical
results for the soft-wall and pure AdS cases for low temperatures.
For higher values of $\widetilde{T}$, the effect of the background
dilaton field is small, and the QNMs are essentially the same as
those of the pure black hole solution \cite{Starinets:2002br,Nunez:2003eq}.
This is related to the fact that the effective potential $V(r_{\ast})$
is a monotonic function without wells both in the intermediate-
and high-temperature regimes.


\TABLE{
\begin{tabular}{cccccc}
\hline
\hline
\vspace{-0.45cm}\\
$\;\;\;\widetilde{T}^{2}\;\;\;$ & $\quad\;\;\;\widetilde{\omega}_{\ss{R}}^{2}\quad\;\;\;$ &
$\quad\;\;\;\widetilde{\omega}_{\ss{I}}^{2}\quad\;\;\;$ & $\quad\;\;\;\Delta_{\ss{R}}
\quad\;\;\;$ & $\quad\;\;\;\Delta_{\ss{I}}\quad\;\;\;$ & $\quad\;\;\;Q\quad\;\;\;$
\\
\hline
100 & 976.020  & 750.874 & 0.00150017 & 0.0023548 & 0.570054\\
25 & 246.247 & 185.138 & 0.0060918 & 0.00923686 & 0.576645\\
4 & 42.1859 & 27.1783 & 0.0410586 & 0.0509827 & 0.622934\\
1 & 13.6137 & 5.45053 & 0.182798 & 0.150013 & 0.790204\\
0.25 & 7.40192 & 0.672991 & 0.744311 & 0.402652 & 1.65820\\
0.04 & 7.72086 & $1.31516\times 10^{-6}$ & 3.45374 & 0.997912
& 1211.47\\
\hline\hline
\end{tabular}
\caption{The frequencies of the fundamental QNM for some selected values
of temperature $\widetilde{T}$, calculated with $q=0$. We also
show the fractional differences in relation to the frequency values in
the absence of dilaton, $\Delta_{\ss{R}}$ and $\Delta_{\ss{I}}$,
and the quality factor of the oscillations, $Q$.}
\label{tabtemperatura}}


\subsubsection{Dispersion relations}

In this section we investigate the momentum dependence of the scalar-field
black hole quasinormal modes in the soft-wall model. In Table
\ref{tabmomentum}, we list some of our numerical results for the real and
(negative of) imaginary parts of the QN frequencies, $\omega_{\ss{R}}(q)$
and $\omega_{\ss{I}}(q)$, for selected values of $\widetilde{T}$ and
two different momenta, $\widetilde{q}=1,\,10\,$. We also compare $\omega_{\ss{R}}(q)$
with the standard relativistic dispersion relation
$\omega=\sqrt{(\omega_{\ss{R}}^{\,\ss{0}})^{2}+q^2}$, where
$\omega_{\ss{R}}^{\,\ss{0}}\equiv\omega_{\ss{R}}(q)|_{_{q=0}}$ can be regarded
as the glueball mass at finite temperature. We introduce the fractional difference
\begin{equation}
\Delta_{q}=\frac{|\omega_{\ss{R}}(q)-\sqrt{(\omega_{\ss{R}}^{\,\ss{0}})^{2}
+q^2}|}{\sqrt{(\omega_{\ss{R}}^{\,\ss{0}})^{2}+q^2}},
\end{equation}
which is a measure of the fractional difference between the numerical values 
obtained for $\omega_{\ss{R}}$ and the relativistic energy-momentum relation
for a massive particle. 

In the ranges of momentum and temperature considered here we have
obtained very small values for $\Delta_{q}$. The maximum value found
for this quantity corresponds approximately to $1.3\,\%$, and appears
for $\widetilde{q}=10$ and $\widetilde{T}^{2}=0.2\,$. The overall behavior
of the dispersion relation $\omega_{\ss{R}}(q)$ is such that $\Delta_{q}$
increases with the momentum and presents an oscillatory dependence in
$\widetilde{T}$ for low temperatures, while increases monotonically for
higher $\widetilde{T}$.

\TABLE{
\begin{tabular}{ccccccc}
\hline\hline
& \multicolumn{3}{c}{$\widetilde{q}=1$} & \multicolumn{3}{c}
{$\widetilde{q}=10$}\\
\cline{2-7}
$\widetilde{T}^{2}$ & $\widetilde{\omega}_{\ss{R}}
$ & $\widetilde{\omega}_{\ss{I}}$ &
$\Delta_{q}$ &
$\widetilde{\omega}_{\ss{R}}$ &
$\widetilde{\omega}_{\ss{I}}$ &
$\Delta_{q}$ \\
\hline
$0.2$ & 2.85042 & 0.644367 & 0.0010234  & 10.4808 & 0.562306 & 0.0127022 \\
$0.15$ & 2.80845 & 0.452717 & 0.000665236  & 10.4244 & 0.449494 & 0.00824847 \\
$0.1$ & 2.80271 & 0.231549 & 0.0024438  & 10.3660 & 0.317525 & 0.00262035 \\
$0.05$ & 2.91009 & 0.0157488 & 0.00267834  & 10.3162 & 0.143700 & 0.00508183 \\
\hline\hline
\end{tabular}
\centering
\caption{Some data for the fundamental QN frequency of the glueball
fluctuations for $\widetilde{q}=1,\,10$ and selected values of temperature.} 
\label{tabmomentum}}

\FIGURE{
\centering\epsfig{file=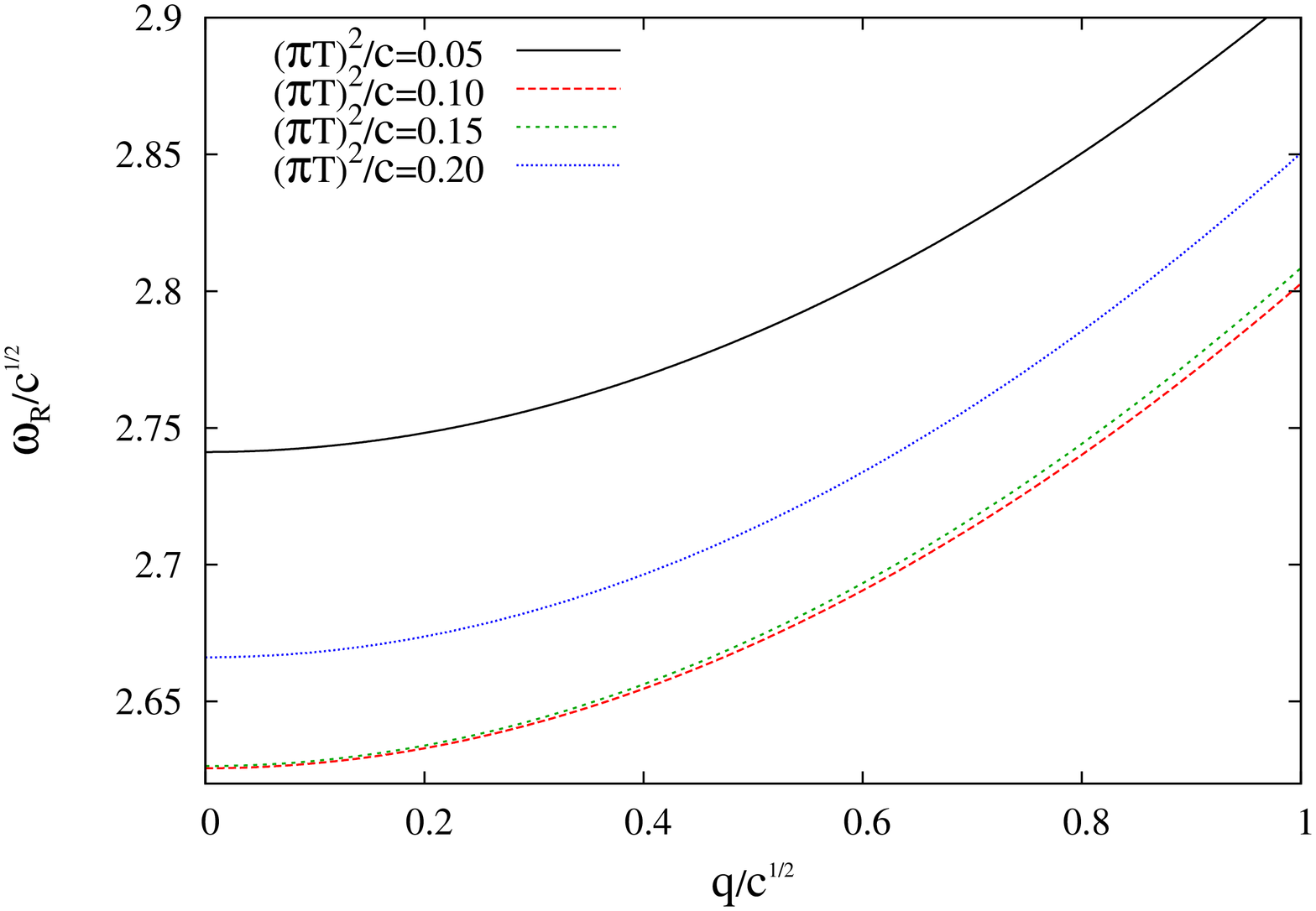,
width=5.15cm, angle=270}
\centering\epsfig{file=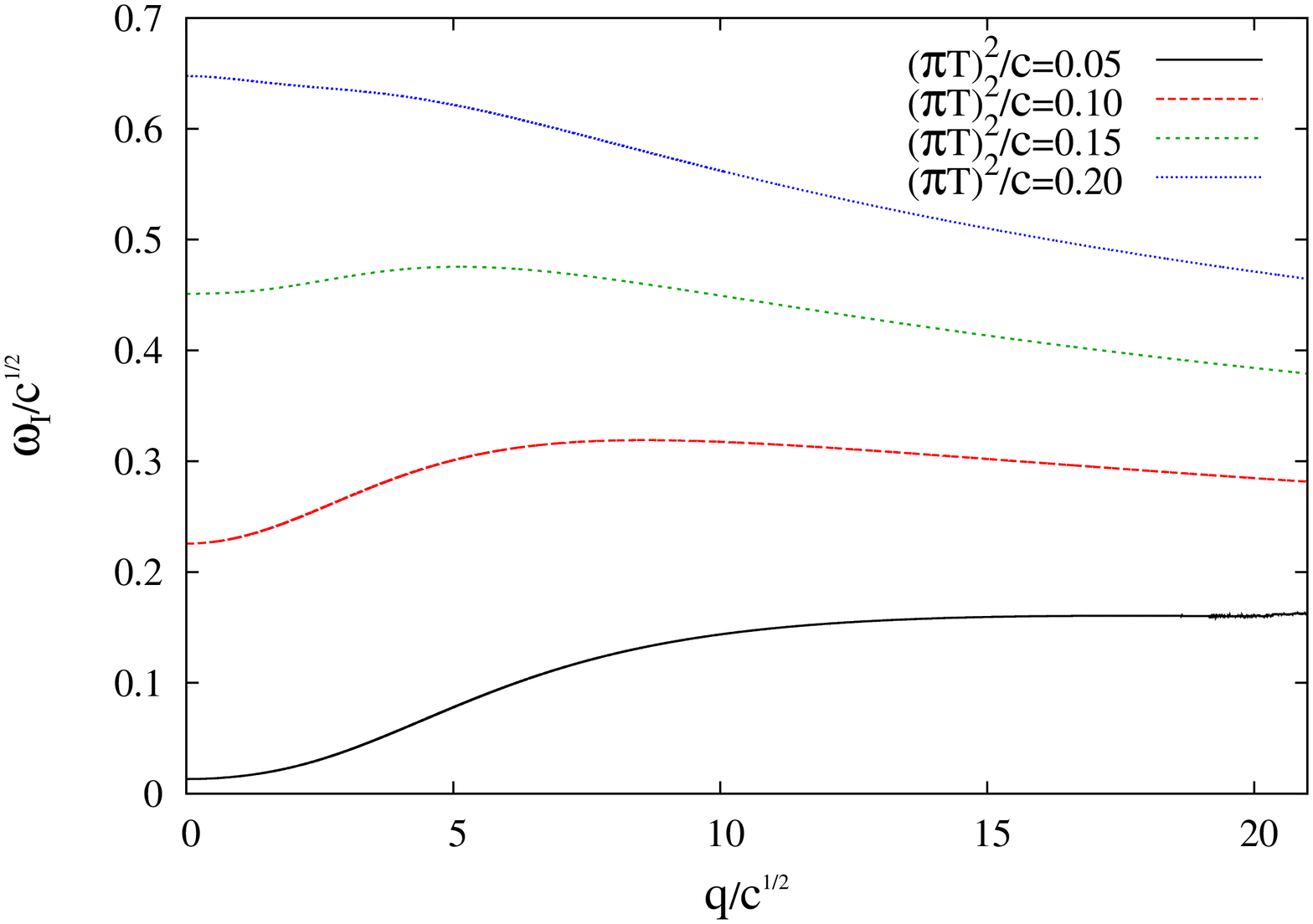,
width=5.15cm, angle=270}
\caption{The dispersion relations of the fundamental
quasinormal mode $n=0$ for some selected values of
temperature.}
\label{QNfreqMomentum}}

\TABLE{
\begin{tabular}{ccccc}
\hline
\hline
\vspace{-0.45cm}\\
$\qquad\widetilde{T}^{2}\qquad$ &
$\qquad\widetilde{q}\qquad$ &
$\qquad\widetilde{\omega}_{\ss{R}}\qquad$ &
$\qquad\widetilde{\omega}_{\ss{I}}^{max}\qquad$ &
$\qquad\tau\qquad$\\
\hline 
0.20 & 0  & 2.66613 & 0.647625 & 1.54410 \\
0.15 & 5.1149 & 5.77971 & 0.475506 & 2.10302 \\
0.10 & 8.5667 & 8.9751 & 0.319018 & 3.13462 \\
\hline\hline
\end{tabular}
\caption{The maximum absolute value of the imaginary part of the fundamental
QN frequency as a function of $\widetilde{q}$, the corresponding values of 
$\widetilde{\omega}_{\ss{R}}$, $\widetilde{T}^2$, and the characteristic damping
time  $\tau=1/\omega_{\ss{I}}$ in units of $c^{-1/2}$. The numerical methods
used here present bad convergence for $\widetilde{T}^2 \lesssim 0.05\,$.}
\label{tabmaximum}}

We present in Fig. \ref{QNfreqMomentum} the variation of the real and
(negative of) imaginary parts of the QN frequency with the momentum
for selected values of the temperature. In the left panel of this
figure, we illustrate the behaviour of the real part of the frequency
in the range $0 \le \widetilde{q} \le 1$. From this picture, one can
see that $\widetilde{\omega}_{\ss{R}}(\widetilde{q})$  increases with
the momentum. In our study we found this behaviour for all temperatures 
investigated.

In the right panel of Fig. \ref{QNfreqMomentum}, we show the dependence 
of (the negative of) the imaginary part of the frequencies with the momentum.
The power series method converged well for momenta in the range 
$0<\widetilde{q}\lesssim 40$ and $\widetilde{T}^{2}=0.2$, but only for
$0<\widetilde{q}\lesssim 20$ when $\widetilde{T}^{2}=0.05\,$.
We found that the absolute value of the imaginary part of the frequency
presents a maximum value, which means a minimum for $\tau=1/\omega_{\ss{I}}$,
for some finite momentum. For temperatures above a critical value
$\widetilde{T}^{2}\approx 0.16$, the maximum value of
$\widetilde{\omega}_{\ss{I}}$ is achieved at $\widetilde{q}=0$.
For temperatures below this critical value, $\widetilde{\omega}_{\ss{I}}$
increases with the momentum for very low  $\widetilde{q}\,$'s, reaches
a maximum value $\widetilde{\omega}_{\ss{I}}^{\,max}$  and then decreases
monotonically. Our numerical results suggest that $\widetilde{\omega}_{\ss{I}}$
vanishes for infinitely large $\widetilde{q}\,$, a behavior similar to
that of the pure AdS black hole case investigated in Refs.
\cite{Festuccia:2008zx,Morgan:2009vg}.

In Table \ref{tabmaximum}, for some selected temperatures $\widetilde{T}$, 
we present the maximum values of $\widetilde{\omega}_{\ss{I}}$ and the corresponding
values of $\widetilde{q}$, $\widetilde{\omega}_{\ss{R}}$, 
and the thermalization timescale $\tau=1/\widetilde{\omega}_{\ss{I}}^{\,max}$.
Below the temperature $\widetilde{T}^{2}=0.05$ the convergence of our numerical
methods gets weak, and we could not determine the value of
$\widetilde{\omega}_{\ss{I}}^{\,max}$ for $\widetilde{T}^{2}\leq 0.05\,$.
In the AdS black hole space without the dilaton, the (negative of the)
imaginary part of the frequency is a decreasing function of the momentum $q$
\cite{Starinets:2002br,Nunez:2003eq}, which is in accordance with the behavior of
$\widetilde{\omega}_{\ss{I}}$ for $\widetilde{T}^{2}>0.16\,$.


\section{Final comments and conclusion}
\label{conclusions}

In this article we investigated the spectrum of scalar glueballs
at finite temperature in the soft-wall model. Such a study was
performed considering a scalar field in an AdS black hole
spacetime in the presence of a dilaton field as the gravity
dual of finite temperature scalar glueballs. This background corresponds 
to the deconfined plasma phase which is thermodynamically favoured
at high temperatures. We considered here the AdS black hole space
for all temperatures. For intermediate temperatures it represents a 
supercooled metastable (positive specific heat) plasma and for low 
temperatures an unstable (negative specific heat) plasma.

We obtained the thermal and momentum dependences of the frequencies of
the black-hole quasinormal modes. For $q=0$, the real part of the QN
frequencies furnishes the glueball masses for low temperatures. 
In the zero temperature limit our results give the glueball spectrum 
found in Ref. \cite{Colangelo:2007pt} for the soft-wall model. 
Surprisingly, for very low temperatures and low fixed momentum, 
these masses decrease with increasing temperature.
After reaching a minimum value, the real part of the QN frequencies
grows monotonically with the temperature, as in the case without
the dilaton field. 

At zero temperature there is an infinite number of stable glueball states
corresponding to delta function peaks in the spectral function.
As the temperature increases, however, the peaks spread, and
for a fixed low temperature and fixed low momentum, there is only
a finite number of sharp peaks, since the width of the peaks
increase with the frequency. The sharp peaks correspond to QNMs
whose complex frequencies have small imaginary parts. These modes
can be interpreted as glueball quasiparticle states with decay rates 
inversely proportional to the imaginary part of the frequencies.
This imaginary part is also proportional to the width of the peak
of the spectral function. 

We studied the dispersion relations for the glueballs in the 
soft wall at finite temperature. This way we found the effect 
of the quasiparticle motion on its lifetime
inside the plasma. The connection of the complex QN frequencies with the mass 
and the decaying time of quasiparticles in the dual plasma 
gives an important tool to explore both sides of the duality using known
elements from the gauge theory and the physics of black holes.

The process of spreading of a peak as the temperature increases is 
interpreted as the melting of the quasiparticles in the thermal bath.
We found that the scalar glueball states melt at temperatures much
lower than the confinement/deconfinement temperature. In the soft-wall
model the phase transition occurs at 
$T=T_c$ with $(\pi T_c)^2/c \approx 2.39$, while we found that,
at $(\pi T)^2/c \approx 0.1$  all the quasiparticles have melted.

Our results indicate that the glueball formation in the black hole phase 
occurs at very low temperatures compared to the
deconfinement temperature. So, from the point of view of the soft-wall model, 
it would be difficult to observe experimentally the contribution of the  black hole phase 
to glueball formation  unless the plasma cools down in a very small time scale, compared 
to that of the thermal phase transition.
Recall that in the soft-wall model there are two different thermal phases, corresponding to
different backgrounds. As shown in ref.\cite{Herzog:2006ra} there is a first order phase 
transition between them, implying that these two phases can coexist.
For  $T < T_c$, the thermal AdS space without a black hole is dominant. 
In this confining background the glueballs are formed at temperatures just below  $T_c$ 
having constant masses and zero widths.

As already mentioned, the soft-wall model is not a solution of the 5D Einstein equations 
since it does not take into account the dilaton backreaction.
There are some recent AdS/QCD models at zero temperature that overcome this problem 
\cite{Gursoy:2007cb,Gursoy:2007er,dePaula:2008fp}.
At finite temperature, the thermodynamics of AdS/QCD models with dilaton backreaction  
was considered in \cite{Gursoy:2008za,Gursoy:2008bu}. 
It would be interesting to extend the analysis of the present work to these kind of models. 
This way one could distinguish the aspects that represent general characteristics of 
quasiparticle excitations in the plasma from those specific to the holographic soft-wall model 
considered here.

After completing this work we noticed the very recent article \cite{Colangelo:2009ra} where glueballs 
and mesons in the soft-wall model at finite temperature are studied, with some overlap 
with this work.


\section*{Acknowledgements}

We thank Vitor Cardoso and Jaqueline Morgan for the help with 
the numerical methods employed in this work. We also thank Marcus
A. C. Torres for stimulating conversations. 
The authors are partially supported by CNPq, Capes and Faperj, 
Brazilian agencies.




\begin{thebibliography}{99}

\bibitem{Maldacena:1997re}
  J.~M.~Maldacena,
  Adv.\ Theor.\ Math.\ Phys.\  {\bf 2}, 231 (1998)
  [Int.\ J.\ Theor.\ Phys.\  {\bf 38}, 1113 (1999)]
  [arXiv:hep-th/9711200].

\bibitem{Witten:1998qj}
  E.~Witten,
  Adv.\ Theor.\ Math.\ Phys.\  {\bf 2}, 253 (1998)
  [arXiv:hep-th/9802150].

\bibitem{Gubser:1998bc}
  S.~S.~Gubser, I.~R.~Klebanov and A.~M.~Polyakov,
  Phys.\ Lett.\  B {\bf 428}, 105 (1998)
  [arXiv:hep-th/9802109].

\bibitem{Witten:1998zw}
  E.~Witten,
  Adv.\ Theor.\ Math.\ Phys.\  {\bf 2}, 505 (1998)
  [arXiv:hep-th/9803131].

\bibitem{Son:2002sd}
  D.~T.~Son and A.~O.~Starinets,
  JHEP {\bf 0209}, 042 (2002)
  [arXiv:hep-th/0205051].

\bibitem{Herzog:2002pc}
  C.~P.~Herzog and D.~T.~Son,
  JHEP {\bf 0303}, 046 (2003)
  [arXiv:hep-th/0212072].

\bibitem{Skenderis:2008dh}
  K.~Skenderis and B.~C.~van Rees,
  Phys.\ Rev.\ Lett.\  {\bf 101}, 081601 (2008)
  [arXiv:0805.0150 [hep-th]].

\bibitem{Skenderis:2008dg}
  K.~Skenderis and B.~C.~van Rees,
  JHEP {\bf 0905}, 085 (2009)
  [arXiv:0812.2909 [hep-th]].

\bibitem{Berti:2009kk}
  E.~Berti, V.~Cardoso and A.~O.~Starinets,
  Class.\ Quant.\ Grav.\  {\bf 26}, 163001 (2009)
  [arXiv:0905.2975 [gr-qc]].

\bibitem{BoschiFilho:2002ta}
  H.~Boschi-Filho and N.~R.~F.~Braga,
  Eur.\ Phys.\ J.\  C {\bf 32}, 529 (2004)
  [arXiv:hep-th/0209080].

\bibitem{BoschiFilho:2002vd}
  H.~Boschi-Filho and N.~R.~F.~Braga,
  JHEP {\bf 0305}, 009 (2003)
  [arXiv:hep-th/0212207].

\bibitem{deTeramond:2005su}
  G.~F.~de Teramond and S.~J.~Brodsky,
  Phys.\ Rev.\ Lett.\  {\bf 94}, 201601 (2005)
  [arXiv:hep-th/0501022].

\bibitem{Erlich:2005qh}
  J.~Erlich, E.~Katz, D.~T.~Son and M.~A.~Stephanov,
Phys. Rev. Lett. {\bf 95} (2005) 261602
[arXiv:hep-ph/0501128].

\bibitem{DaRold:2005zs}
  L.~Da Rold and A.~Pomarol,
 Nucl.\ Phys.\  {\bf B 721} (2005) 79
[arXiv:hep-ph/0501218].

\bibitem{BoschiFilho:2005yh}
  H.~Boschi-Filho, N.~R.~F.~Braga and H.~L.~Carrion,
  Phys.\ Rev.\  D {\bf 73}, 047901 (2006)
  [arXiv:hep-th/0507063].

\bibitem{BoschiFilho:2006pe}
  H.~Boschi-Filho, N.~R.~F.~Braga and C.~N.~Ferreira,
  Phys.\ Rev.\  D {\bf 74}, 086001 (2006)
  [arXiv:hep-th/0607038].

\bibitem{BoschiFilho:2005mw}
  H.~Boschi-Filho, N.~R.~F.~Braga and C.~N.~Ferreira,
  Phys.\ Rev.\  D {\bf 73}, 106006 (2006)
  [Erratum-ibid.\  D {\bf 74}, 089903 (2006)]
  [arXiv:hep-th/0512295].

\bibitem{Karch:2006pv}
  A.~Karch, E.~Katz, D.~T.~Son and M.~A.~Stephanov,
  Phys.\ Rev.\  D {\bf 74}, 015005 (2006)
  [arXiv:hep-ph/0602229].

\bibitem{Colangelo:2007pt}
  P.~Colangelo, F.~De Fazio, F.~Jugeau and S.~Nicotri,
  Phys.\ Lett.\  B {\bf 652}, 73 (2007)
  [arXiv:hep-ph/0703316].

\bibitem{Colangelo:2008us}
  P.~Colangelo, F.~De Fazio, F.~Giannuzzi, F.~Jugeau and S.~Nicotri,
  Phys.\ Rev.\  D {\bf 78}, 055009 (2008)
  [arXiv:0807.1054 [hep-ph]].

\bibitem{Andreev:2006eh}
  O.~Andreev and V.~I.~Zakharov,
  Phys.\ Lett.\  B {\bf 645}, 437 (2007)
  [arXiv:hep-ph/0607026].

\bibitem{Herzog:2006ra}
  C.~P.~Herzog,
  Phys.\ Rev.\ Lett.\  {\bf 98}, 091601 (2007)
  [arXiv:hep-th/0608151].

\bibitem{Kajantie:2006hv}
  K.~Kajantie, T.~Tahkokallio and J.~T.~Yee,
  JHEP {\bf 0701}, 019 (2007)
  [arXiv:hep-ph/0609254].

\bibitem{BallonBayona:2007vp}
  C.~A.~Ballon Bayona, H.~Boschi-Filho, N.~R.~F.~Braga and L.~A.~Pando Zayas,
  Phys.\ Rev.\  D {\bf 77}, 046002 (2008)
  [arXiv:0705.1529 [hep-th]].

\bibitem{Hawking:1982dh}
  S.~W.~Hawking and D.~N.~Page,
  Commun.\ Math.\ Phys.\  {\bf 87}, 577 (1983).

\bibitem{Myers:2007we}
  R.~C.~Myers, A.~O.~Starinets and R.~M.~Thomson,
  JHEP {\bf 0711}, 091 (2007)
  [arXiv:0706.0162 [hep-th]].

\bibitem{Horowitz:1999jd}
  G.~T.~Horowitz and V.~E.~Hubeny,
  Phys.\ Rev.\  D {\bf 62}, 024027 (2000)
  [arXiv:hep-th/9909056].

\bibitem{Berti:2009wx}
  E.~Berti, V.~Cardoso and P.~Pani,
  Phys.\ Rev.\  D {\bf 79}, 101501 (2009)
  [arXiv:0903.5311 [gr-qc]].


\bibitem{Forkel:2007ru}
  H.~Forkel,
  Phys.\ Rev.\  D {\bf 78}, 025001 (2008)
  [arXiv:0711.1179 [hep-ph]].

\bibitem{Mathieu:2008me}
  V.~Mathieu, N.~Kochelev and V.~Vento,
  Int.\ J.\ Mod.\ Phys.\  E {\bf 18}, 1 (2009)
  [arXiv:0810.4453 [hep-ph]].


\bibitem{Gibbons:2002pq}
  G.~Gibbons and S.~A.~Hartnoll,
  Phys.\ Rev.\  D {\bf 66}, 064024 (2002)
  [arXiv:hep-th/0206202].

\bibitem{Kodama:2003jz}
  H.~Kodama and A.~Ishibashi,
  Prog.\ Theor.\ Phys.\  {\bf 110}, 701 (2003)
  [arXiv:hep-th/0305147].

\bibitem{Morgan:2009pn}
  J.~Morgan, V.~Cardoso, A.~S.~Miranda, C.~Molina and V.~T.~Zanchin,
  JHEP {\bf 0909}, 117 (2009)
  [arXiv:0907.5011 [hep-th]].

\bibitem{Bianchi:2001kw}
  M.~Bianchi, D.~Z.~Freedman and K.~Skenderis,
  Nucl.\ Phys.\  B {\bf 631}, 159 (2002)
  [arXiv:hep-th/0112119].

\bibitem{Fujita:2009wc}
  M.~Fujita, K.~Fukushima, T.~Misumi and M.~Murata,
  Phys.\ Rev.\  D {\bf 80}, 035001 (2009)
  [arXiv:0903.2316 [hep-ph]].

\bibitem{Kovtun:2006pf}
  P.~Kovtun and A.~Starinets,
  Phys.\ Rev.\ Lett.\  {\bf 96}, 131601 (2006)
  [arXiv:hep-th/0602059].

\bibitem{Teaney:2006nc}
  D.~Teaney,
  Phys.\ Rev.\  D {\bf 74}, 045025 (2006)
  [arXiv:hep-ph/0602044].

\bibitem{Myers:2008cj}
  R.~C.~Myers and A.~Sinha,
  JHEP {\bf 0806}, 052 (2008)
  [arXiv:0804.2168 [hep-th]].

\bibitem{Myers:2008me}
  R.~C.~Myers and M.~C.~Wapler,
  JHEP {\bf 0812}, 115 (2008)
  [arXiv:0811.0480 [hep-th]].

 
\bibitem{Colangelo:2007if}
  P.~Colangelo, F.~De Fazio, F.~Jugeau and S.~Nicotri,
  Int.\ J.\ Mod.\ Phys.\  A {\bf 24}, 4177 (2009)
  [arXiv:0711.4747 [hep-ph]].


\bibitem{Kokkotas:1999bd}
  K.~D.~Kokkotas and B.~G.~Schmidt,
  Living Rev.\ Rel.\  {\bf 2}, 2 (1999)
  [arXiv:gr-qc/9909058].

\bibitem{Nollert:1999ji}
  H.~P.~Nollert,
  Class.\ Quant.\ Grav.\  {\bf 16}, R159 (1999).

\bibitem{Wang:2000dt}
  B.~Wang, C.~Molina and E.~Abdalla,
  Phys.\ Rev.\  D {\bf 63}, 084001 (2001)
  [arXiv:hep-th/0005143].

\bibitem{Cardoso:2001hn}
  V.~Cardoso and J.~P.~S.~Lemos,
  Phys.\ Rev.\  D {\bf 63}, 124015 (2001)
  [arXiv:gr-qc/0101052].

\bibitem{Cardoso:2001bb}
  V.~Cardoso and J.~P.~S.~Lemos,
  Phys.\ Rev.\  D {\bf 64}, 084017 (2001)
  [arXiv:gr-qc/0105103].

\bibitem{Starinets:2002br}
  A.~O.~Starinets,
  Phys.\ Rev.\  D {\bf 66}, 124013 (2002)
  [arXiv:hep-th/0207133].

\bibitem{Nunez:2003eq}
  A.~Nunez and A.~O.~Starinets,
  Phys.\ Rev.\  D {\bf 67}, 124013 (2003)
  [arXiv:hep-th/0302026].

\bibitem{Kovtun:2005ev}
  P.~K.~Kovtun and A.~O.~Starinets,
  Phys.\ Rev.\  D {\bf 72}, 086009 (2005)
  [arXiv:hep-th/0506184].

\bibitem{Maeda:2005cr}
  K.~Maeda, M.~Natsuume and T.~Okamura,
  Phys.\ Rev.\  D {\bf 72}, 086012 (2005)
  [arXiv:hep-th/0509079].

\bibitem{Siopsis:2004as}
  G.~Siopsis,
  Nucl.\ Phys.\  B {\bf 715}, 483 (2005)
  [arXiv:hep-th/0407157].

\bibitem{Miranda:2005qx}
  A.~S.~Miranda and V.~T.~Zanchin,
  Phys.\ Rev.\  D {\bf 73}, 064034 (2006)
  [arXiv:gr-qc/0510066].

\bibitem{Zhang:2006bc}
  Y.~Zhang and J.~L.~Jing,
  Int.\ J.\ Mod.\ Phys.\  D {\bf 15}, 905 (2006).

\bibitem{Hoyos:2006gb}
  C.~Hoyos-Badajoz, K.~Landsteiner and S.~Montero,
  JHEP {\bf 0704}, 031 (2007)
  [arXiv:hep-th/0612169].

\bibitem{Miranda:2007bv}
  A.~S.~Miranda and V.~T.~Zanchin,
  Int.\ J.\ Mod.\ Phys.\  D {\bf 16}, 421 (2007).

\bibitem{Amado:2008hw}
  I.~Amado and C.~Hoyos-Badajoz,
  JHEP {\bf 0809}, 118 (2008)
  [arXiv:0807.2337 [hep-th]].

\bibitem{arfken} 
G. B. Arfken and H. J. Weber, {\it Mathematical Methods for Physicists}, 
4th. ed. San Diego, CA: Academic Press, 1995. 

\bibitem{Thorne:1969}
  K.~S.~Thorne,
  Astrophys.\ J.\  {\bf 158}, 1 (1969).

\bibitem{ChandrasekharFerrari}
  S.~Chandrasekhar and V.~Ferrari,
  Proc.\ Roy.\ Soc.\ Lond.\  A {\bf 434}, 449 (1991).

\bibitem{Chandrasekhar:1992ey}
  S.~Chandrasekhar and V.~Ferrari,
  Proc.\ Roy.\ Soc.\ Lond.\  A {\bf 437}, 133 (1992).

\bibitem{Festuccia:2008zx}
  G.~Festuccia and H.~Liu,
  arXiv:0811.1033 [gr-qc].

\bibitem{Morgan:2009vg}
  J.~Morgan, V.~Cardoso, A.~S.~Miranda, C.~Molina and V.~T.~Zanchin,
  Phys.\ Rev.\  D {\bf 80}, 024024 (2009)
  [arXiv:0906.0064 [hep-th]].

\bibitem{Gursoy:2007cb}
  U.~Gursoy and E.~Kiritsis,
  JHEP {\bf 0802}, 032 (2008)
  [arXiv:0707.1324 [hep-th]].

\bibitem{Gursoy:2007er}
  U.~Gursoy, E.~Kiritsis and F.~Nitti,
  JHEP {\bf 0802}, 019 (2008)
  [arXiv:0707.1349 [hep-th]].

\bibitem{dePaula:2008fp}
  W.~de Paula, T.~Frederico, H.~Forkel and M.~Beyer,
  Phys.\ Rev.\  D {\bf 79}, 075019 (2009)
  [arXiv:0806.3830 [hep-ph]].
  
\bibitem{Gursoy:2008za}
  U.~Gursoy, E.~Kiritsis, L.~Mazzanti and F.~Nitti,
  JHEP {\bf 0905}, 033 (2009)
  [arXiv:0812.0792 [hep-th]].
    
\bibitem{Gursoy:2008bu}
  U.~Gursoy, E.~Kiritsis, L.~Mazzanti and F.~Nitti,
  Phys.\ Rev.\ Lett.\  {\bf 101}, 181601 (2008)
  [arXiv:0804.0899 [hep-th]].
  
\bibitem{Colangelo:2009ra}
  P.~Colangelo, F.~Giannuzzi and S.~Nicotri,
  arXiv:0909.1534 [hep-ph].
  
  
   
\end{thebibliography}
\end{document}